\documentclass[aps,prc,preprint,groupedaddress]{revtex4-1}
\usepackage{graphicx}
\usepackage{amssymb}
\usepackage{amsmath}
\usepackage{color}
\usepackage{setspace}
\usepackage{amsfonts}
\newcommand{\cgc}[6]{ \left\langle {#1}\,{#2}\,{#3}\,{#4} \middle\vert {#5}\,{#6} \right\rangle}

\newcommand{\braket}[2]{\left\langle {#1} \middle\vert {#2} \right\rangle}

\newcommand{\braoket}[3]{\left\langle {#1} \middle\vert {#2} \middle\vert {#3} \right\rangle}

\makeatletter
	\def\erf{\mathop{\operator@font erf}\nolimits}
	\def\erfc{\mathop{\operator@font erfc}\nolimits}
	\def\Erf{\mathop{\operator@font Erf}\nolimits}
	\def\Shi{\mathop{\operator@font Shi}\nolimits}
	\def\Chi{\mathop{\operator@font Chi}\nolimits}
	\def\Ei{\mathop{\operator@font Ei}\nolimits}
	\def\cosec{\mathop{\operator@font cosec}\nolimits}
	\def\sech{\mathop{\operator@font sech}\nolimits}
	\def\cosech{\mathop{\operator@font cosech}\nolimits}
	\newcommand\hypgeo[2]{{}_{#1}{\operator@font F}_{#2}}
\makeatother
\begin{document}

%\preprint{}

\title{Neutron spin dynamics in polarized targets}



\author{Vladimir Gudkov}
\email[]{gudkov@sc.edu}
\affiliation{Department of Physics and Astronomy, University of South Carolina, Columbia, South Carolina 29208, USA}

\author{Hirohiko M. Shimizu}
\email[]{shimizu@phi.phys.nagoya-u.ac.jp}
\affiliation{Department of Physics, Nagoya University, Nagoya 464-8602, Japan}


\date{\today}

\begin{abstract}
We present neutron elastic scattering amplitude for arbitrary polarized target in irreducible spherical tensor representation.
The general approach for the description of neutron spin dynamics for the propagation trough the medium with an arbitrary polarization is discussed in a relation to the search for time reversal invariance violation in neutron scattering.
\end{abstract}

%\pacs{24.80.+y,  11.30.Er,    25.40.Dn}
%\keywords{ }

\maketitle

%\section{Introduction}

\section{Introduction}

With the opportunity to measure
time reversal invariance violating (TRIV)
  effects in nuclear reactions by the transmission of polarized neutrons through a polarized target\cite{Kabir:1982tp,Stodolsky:1982tp,Bunakov:1982is,Gudkov:1991qg,Bowman:2014fca} it is important to have a complete description of the propagation of neutron spin through arbitrary polarized target. These TRIV effects are proportional to vector polarization of nuclear target. However,  in a general case, the polarization of a target with spin $I$ requires  $2I$ tensor momenta \cite{Polarization:1971,Varshalovich,NuclPolar:1961}. Therefore,  only for $I=1/2$ it is sufficient to consider vector polarization (the first rank tensor) for complete description of the target polarization.
      In spite of the fact that the recent proposals for searches TRIV in neutron-nucleus scattering (see, for example, Ref. \cite{Bowman:2014fca} and references therein)  demonstrated the existence of a class of experiments that are free from false asymmetries,
 to design the experiment and to control the possible systematic effects one has to have a detailed description of neutron spin dynamics in  targets with arbitrary polarization. The propagation of polarized neutrons through polarized target  in relation to TRIV experiments have been studied in many papers (see, for example
 \cite{Bunakov:1987rb,Stodolsky:1985ic,Kabir:1988ma,Kabir:1989pc,Skoy:1996,Bowman:2014fca,Gudkov:2017sye} and references therein), however these studies have been done with the focus on the case of vector polarized target.
 However, even for 100\% vector polarized target with spin $I>1/2$, the higher rank tensor polarizations may coexist and to be rather large. Therefore, even  if these higher order polarizations cannot mimic TRIV effects, they can change the neutron spin dynamics, which can lead to a suppression of TRIV observables.   Thus, the open and important question for design new experiments and for the future data analysis question is how these high rank tensor polarizations affect neutron spin dynamics inside the polarized target.

 In this paper we give a systematic approach for the description of low energy neutron spin propagation in the target with arbitrary polarization using irreducible spherical tensor representation for target polarizations. Also, as  examples, we present the complete expressions for neutron scattering amplitude in irreducible spherical tensor representation   for the case of a nuclear target with spin $I=7/2$, and the detailed analysis of neutron scattering on $^{139}La$. Finally, we apply the developed approach for a general analysis of neutron spin rotation in a target with arbitrary polarization.

\section{Scattering amplitude in irreducible spherical tensor representation.}

 A general expression for the  forward elastic scattering amplitude with $\mu$ and $M$,  projections of neutron ($s =1/2$) and target spins on a quantization axis, can be written as
\begin{eqnarray}\label{genAmp}
 \nonumber
 f_{MM^{\prime}}  & =\frac{i \pi }{2k}& \sum_{Jll^{\prime}  S m_s S^{\prime} m^{\prime}_s}Y_{L m_L}(\theta ,\phi)
  	\left\langle s \mu^{\prime} I M^\prime \middle\vert S^{\prime} m^{\prime}_s \right\rangle
	\left\langle S m_s \middle\vert s \mu I M \right\rangle\\
  &\times&
   \nonumber
  	\left\langle S^{\prime} l^{\prime} \alpha^{\prime} \middle\vert R^J \middle\vert S l \alpha\right\rangle
	(-1)^{J+S^{\prime}+l^{\prime}+l}(2J+1)\sqrt{\frac{(2l+1)(2l^{\prime}+1)}{4\pi (2S+1)}} \\
	 &\times&
\left\langle l 0 l^{\prime} 0 \middle\vert L 0 \right\rangle
\left\langle L m_L S^{\prime} m^{\prime}_s \middle\vert S m_s \right\rangle
 \begin{Bmatrix}
   l^{\prime} & l & L \\
   S & S^{\prime} & J
  \end{Bmatrix},
 \end{eqnarray}
where  primed parameters correspond to outgoing channel, and the angles $\theta$ and $\phi$ describe a direction of neutron momentum $\vec{k}$. The matrix $\hat{R}$ is related to the scattering matrix  $\hat{\mathbb{S}}$ as $\hat{R}=\hat{1}-\hat{\mathbb{S}}$ which  in the integral of motion representation \cite{Baldin:1961} is
\begin{equation}\label{Smat}
\left\langle S^{\prime} l^{\prime} \alpha^{\prime} \middle\vert \mathbb{S}^J \middle\vert S l \alpha \right\rangle
\delta_{JJ^{\prime}}\delta_{MM^{\prime}}\delta(E^{\prime} -E) ,
\end{equation}
where $J$ and $M$ are the total spin and its projection, $S$ is the channel spin, $l$ is the orbital momentum, and $\alpha$ represents the other internal quantum numbers.

 Then, for arbitrary target polarization described by polarization density matrix $\rho_{MM^\prime}$ the scattering amplitude can be calculated as
 \begin{equation}\label{tensAmp}
   f=Tr(f_{MM^\prime}\rho_{MM^\prime}).
 \end{equation}
To describe polarization of  tensor polarized target, it is convenient to use the expansion \cite{Varshalovich} of the density polarization matrix in terms of the statistical tensors $t_{q \kappa}$
 \begin{equation}\label{denMart}
   \rho_{MM^\prime}=\sum_{q \kappa}\sqrt{\frac{2q+1}{2I+1}}\left\langle I M q \kappa \middle\vert I M^\prime \right\rangle t_{q \kappa}.
 \end{equation}
   In this expression each tensor $t_{q \kappa}$ corresponds to tensor polarization of the target of a rank $q$, thus for the case of unpolarized target all statistical tensors vanish except $t_{0 0}$.

For the choice of the spin direction of the target along the quantization axis $z$,  the elastic scattering amplitude (\ref{tensAmp}) can be presented as an expansion in terms of  spherical-tensor polarizations $t_{q0}$ with a corresponding weight of $w_q$:
 \begin{eqnarray}\label{tensAmp}
 \nonumber
 f  & = \frac{i\pi }{2k}& \sum^{2I}_{q=0}w_q t_{q0} \sqrt{\frac{2q+1}{2I+1}} \left[ \sum_{Jll^{\prime}  S m_s S^{\prime} m^{\prime}_s}Y_{L m_L}(\theta ,\phi)
  	\left\langle s \mu^{\prime} I M \middle\vert S^{\prime} m^{\prime}_s \right\rangle \right.
	\left\langle S m_s \middle\vert s \mu I M \right\rangle  \\
  &\times&
   \nonumber
   \left\langle I M q 0 \middle\vert I M \right\rangle
  	\left\langle S^{\prime} l^{\prime} \alpha^{\prime} \middle\vert R^J \middle\vert S l \alpha\right\rangle
	(-1)^{J+S^{\prime}+l^{\prime}+l}(2J+1)\sqrt{\frac{(2l+1)(2l^{\prime}+1)}{4\pi (2S+1)}} \\
	 &\times& \left.
\left\langle l 0 l^{\prime} 0 \middle\vert L 0 \right\rangle
\left\langle L m_L S^{\prime} m^{\prime}_s \middle\vert S m_s \right\rangle
 \begin{Bmatrix}
   l^{\prime} & l & L \\
   S & S^{\prime} & J
  \end{Bmatrix} \right].
 \end{eqnarray}
For a description of neutron polarization
 it is convenient to introduce a function:
\begin{eqnarray}\label{Pfun}
   & & N(\tilde{x},\tilde{y},S, S^{\prime},M , M^{\prime}) = \frac{1}{|\tilde{x}|^2+|\tilde{y}|^2} \\
   \nonumber
   &\times&\left( |\tilde{x}|^2 \left\langle \frac{1}{2}\frac{1}{2} I M \middle\vert S M+\frac{1}{2} \right\rangle
  \left\langle \frac{1}{2} \frac{1}{2} I M^{\prime} \middle\vert S^{\prime} M  +\frac{1}{2} \right\rangle
   \left\langle L 0 S^{\prime} M^{\prime}  +\frac{1}{2} \middle\vert S M  +\frac{1}{2} \right\rangle  \delta_{m_L,0} \right.   \\
     &+& \tilde{x}\tilde{y}^*
  \left\langle \frac{1}{2}-\frac{1}{2} I M \middle\vert S M-\frac{1}{2} \right\rangle
  \left\langle \frac{1}{2} \frac{1}{2} I M^{\prime} \middle\vert S^{\prime} M^{\prime}  +\frac{1}{2} \right\rangle
   \left\langle L 1 S^{\prime} M^{\prime}  -\frac{1}{2} \middle\vert S M  +\frac{1}{2} \right\rangle  \delta_{m_L,1} \\
   \nonumber
   &+& \tilde{x}^*\tilde{y}
  \left\langle \frac{1}{2}-\frac{1}{2} I M \middle\vert S M-\frac{1}{2} \right\rangle
  \left\langle \frac{1}{2} \frac{1}{2} I M^{\prime} \middle\vert S^{\prime} M  +\frac{1}{2} \right\rangle
   \left\langle L -1 S^{\prime} M^{\prime}  +\frac{1}{2} \middle\vert S M  -\frac{1}{2} \right\rangle  \delta_{m_L,-1} \\
   \nonumber
   &+& \left. |\tilde{y}|^2 \left\langle \frac{1}{2}-\frac{1}{2} I M \middle\vert S M-\frac{1}{2} \right\rangle
  \left\langle \frac{1}{2}- \frac{1}{2} I M^{\prime} \middle\vert S^{\prime} M -\frac{1}{2} \right\rangle
   \left\langle L 0 S^{\prime} M^{\prime}  -\frac{1}{2} \middle\vert S M  -\frac{1}{2} \right\rangle  \delta_{m_L,0}      \right) ,
\end{eqnarray}
where $\tilde{x}$ and $\tilde{y}$ are the components of neutron spinor $\binom{\tilde{x}}{\tilde{y}}$. To describe an arbitrary orientation of neutron spin in spherical coordinates, one can choose
\begin{eqnarray} \label{spinSph}
 \nonumber
  \tilde{x} &=& \cos (\beta /2)e^{-i\alpha /2} \\
  \tilde{y} &=& \sin (\beta /2)e^{i\alpha /2} ,
\end{eqnarray}
where $\beta$ and $\alpha$ are polar and azimuth angles for spin direction relative to the quantization axis $z$.
Then, the elastic scattering amplitude for polarized neutrons and for  spherical-tensor polarization $P^I_q$ (as defined in eq.(\ref{defpol}))  of the target is
 \begin{eqnarray}\label{tensAmpNeutr}
 \nonumber
 f (\tilde{x},\tilde{y})  & = \frac{i\pi }{2k}& \sum^{2I}_{q=0}\frac{P^I_q}{c^I_q} \widetilde{\tau}_{q 0} \sqrt{2q+1} \left[ \sum_{JMll^{\prime}  S  S^{\prime} }Y_{L m_L}(\theta ,\phi )
  	N(\tilde{x},\tilde{y},S, S^{\prime},M , M)\right. \\
  &\times&
   \left\langle I M q 0 \middle\vert I M \right\rangle
    \nonumber
         	\left\langle S^{\prime} l^{\prime} \alpha^{\prime} \middle\vert R^J \middle\vert S l \alpha\right\rangle
	(-1)^{J+S^{\prime}+l^{\prime}+l}(2J+1)\sqrt{\frac{(2l+1)(2l^{\prime}+1)}{4\pi (2S+1)}} \\
	 &\times& \left.
\left\langle l 0 l^{\prime} 0 \middle\vert L 0 \right\rangle
 \begin{Bmatrix}
   l^{\prime} & l & L \\
   S & S^{\prime} & J
  \end{Bmatrix} \right] = \sum^{2I}_{q=0}P^I_q f_q,
 \end{eqnarray}
with $c^I_q$ defined by eq.(\ref{polcoeff}).

It should be noted that usually in nuclear physics  spherical tensors (which we define as $\widetilde{\tau}_{\kappa q}$ in appendix \ref{tensor} ) have different normalization \cite{Lakin,Polarization:1971,NuclPolar:1961} compare to $t_{qk}$ in eq.(\ref{denMart}), defined in \cite{Varshalovich}.
Therefore for the sake of convenience we will present further expressions for amplitudes and angular coefficients in terms of expansions in $\widetilde{\tau}_{q0}$.

  To  describe  the amplitude for arbitrary target's and neutron's spin orientations, we use a convention that direction of the target spin is alway parallel to the axis $z$, and the neutron momentum   belongs to the  $y$-$z$ plane, which imply that the angle  $\phi =\pi/2$ and the neutron momentum  direction is described by the angle $\theta$. Then, assuming that $\vec{\sigma}$, $\vec{I}$, and $\vec{k}$ are unit vectors in the corresponding directions, we can write the following relations
\begin{eqnarray}\label{angRel}
 \nonumber
 (\vec{k}\cdot\vec{I}) &=& \cos \theta , \\
  \nonumber
   (\vec{\sigma}\cdot\vec{I}) &=& \cos \beta ,\\
  \nonumber
  (\vec{\sigma}\cdot\vec{k}) &=& \cos \theta \cos \beta + \sin \theta \sin \beta \sin \alpha ,\\
  \nonumber
  [\vec{k}\times\vec{I}]  &=& \sin \theta , \\
   (\vec{\sigma}\cdot[\vec{k}\times\vec{I}]) &=& \cos \alpha \sin \theta \sin \beta .
\end{eqnarray}
Using these relations one can expand  the scattering amplitude (\ref{tensAmpNeutr}) in terms of irreducible tensors constructed from  the products of
  neutron spin $\vec{\sigma}$,   target  spin $\vec{I}$, and   neutron momentum $ \vec{k}$.  In general, only the first order of neutron spin, and the powers of the target spin up to the value of $(2I)$ can contribute in this expansion,  the power of neutron momentum is not bounded. However, since we consider scattering of low energy neutrons with only contributions from $s$-, and $p$- partial waves (resonances), we can restrict the power of  $ \vec{k}$ by a factor of two. Then,  the amplitude ((\ref{tensAmpNeutr})) can be written as:
\begin{eqnarray}\label{gamp}
\nonumber
  f &=& A^{\prime}+B^{\prime}(\vec{\sigma}\cdot \vec{I})+C^{\prime}(\vec{\sigma}\cdot \vec{k})+D^{\prime}(\vec{\sigma}\cdot [\vec{k}\times \vec{I}])+ H^{\prime}(\vec{k}\cdot \vec{I}) + K^{\prime}(\vec{\sigma}\cdot \vec{k})(\vec{k}\cdot \vec{I}) \\
  \nonumber
  &+&E^{\prime}\left( (\vec{k}\cdot \vec{I})(\vec{k}\cdot \vec{I})-\frac{1}{3}(\vec{k}\cdot \vec{k})(\vec{I}\cdot \vec{I})\right)
  +  F^{\prime}\left( (\vec{\sigma}\cdot \vec{I})(\vec{k}\cdot \vec{I})-\frac{1}{3}(\vec{\sigma}\cdot \vec{k})(\vec{I}\cdot \vec{I})\right)\\
   &+&G^{\prime}(\vec{\sigma}\cdot [\vec{k}\times \vec{I}])(\vec{k}\cdot \vec{I})+B_3^{\prime}(\vec{\sigma}\cdot \vec{I})\left( (\vec{k}\cdot \vec{I})(\vec{k}\cdot \vec{I})-\frac{1}{3}(\vec{k}\cdot \vec{k})(\vec{I}\cdot \vec{I})\right) +...,
\end{eqnarray}
where the first line contains target spin independent terms ($A^{\prime}$, $C^{\prime}$) and terms proportional to the vector polarization of the target ($B^{\prime}$, $D^{\prime}$, $H^{\prime}$, $K^{\prime}$), while the second and the third lines contain  $E^{\prime}$, $F^{\prime}$, and $G^{\prime}$ terms, which are  proportional to to the tensor polarization of the second rank, and the term $B_3^{\prime}$, which is proportional to the third rank of the target polarization.  The terms $C^{\prime}$, $D^{\prime}$, $H^{\prime}$  and $F^{\prime}$ represent P-odd of the amplitude, and  $D^{\prime}$ and $G^{\prime}$ terms  violate time reversal invariance. (The expressions for these coefficients for  specific values of the target spins $I$ and their projections $M$ on the axis $z$ are given in appendix \ref{Mappendix}.)

For parity conserving parts of the  amplitude (\ref{tensAmpNeutr}) and, as a consequence, of the amplitude (\ref{gamp}),
the  matrix elements  for slow neutrons  can be written in the Breit-Wigner resonance approximation  as \cite{Bunakov:1982is}
\begin{eqnarray}\label{BWamp}
F_{S'Sl}^{J}
&\equiv&
\left\langle S^{\prime} lK \middle\vert R^{J} \middle\vert S lK \right\rangle
\nonumber\\
&=& \sum_K
i \frac
{\sqrt{\Gamma^{\rm n}_{l_K}(S^{\prime}_K)}\sqrt{\Gamma^{\rm n}_{l_K}(S_K)}}
{E-E_K+i\Gamma_K/2}
e^{i(\delta_{l_K}(S^{\prime}_K)+\delta_{l_K}(S_K))}
-2ie^{i\delta_{l_K}(S_KS^{\prime}_K)}\sin\delta_{l_K}(S_K S^{\prime}_K)
\nonumber\\
\label{eq:1}
\end{eqnarray}
where $E_K$, $\Gamma_K$, and $\Gamma^n_{l_K}$ are the energy, the total width, and the partial neutron width of the $K$-th nuclear compound resonance, $E$ is the neutron energy, and $\delta_{l_K}$ is the potential scattering phase shift.
For $p$-wave resonances we keep only the resonance term, because for low energy neutrons $\delta_l\sim (kR_0)^{2l+1}$ (where $R_0$ is nucleus radius), and, as a consequence, the contribution from $p$-wave potential scattering is negligible.

The matrix elements for PV and TRIV interactions for slow neutrons can be written  in the Breit-Wigner resonance approximation with one $s$-resonance and one $p$-resonance as \cite{Bunakov:1982is,Gudkov:1990tb}
\begin{eqnarray}\label{BWampSP}
\left(F_{S'l^{\prime}Sl}^{J}\right)
&\equiv&
\left\langle S^{\prime} l^{\prime} \middle\vert R^{J} \middle\vert S l \right\rangle
\nonumber\\
&=&
 \frac
{\sqrt{\Gamma^{\rm n}_{l^{\prime}}(S^{\prime})}(-iv+w)\sqrt{\Gamma^{\rm n}_{l}(S)}}
{(E-E_l+i\Gamma_l/2)(E-E_{l^{\prime}}+i\Gamma_{l^{\prime}}/2)}
e^{i(\delta_{l^{\prime}}(S^{\prime})+\delta_{l}(S))}
\nonumber\\
\label{eq:1}
\end{eqnarray}
where $l\neq l^{\prime}$, and  $v$ and $w$ are real and imaginary parts  of the matrix elements for PV and TRIV mixing between
$s$- and $p$-wave compound resonances
\begin{equation}
 v+iw = -<{\phi_s}|V_{\not{P}}+V_{\not{P}\not{T}}|{\phi_p}>
  \label{eq:me}
\end{equation}
due to   $V_{\not{P}}$ (PV) and $V_{\not{P}\not{T}}$ (TRIV) interactions.

In general, the matrix element   eq.(\ref{BWampSP}) has a sum over a number of close resonances, similar to the sum in eq.(\ref{BWamp}). However,  we are usually interested in a description of symmetry violating effects in the vicinity of  $p$-resonances. In that case, only a contribution from that particular $p$-resonance is important, therefore  we can use two resonance approximation (\ref{BWampSP}) resulted from a mixture of the nearest  $s$- and $p$-resonances. It should be noted, that in  general  $p$-resonance can be mixed with two or more $s$-wave resonances. In that case, the eq.(\ref{BWampSP}) should be modified to the sum of (\ref{BWampSP})amplitudes over all mixing $s$-resonances.
 Fortunately, the  two resonance approximation has been proved to be good enough to describe practically all observed  PV effects in neutron scattering (see, \cite{Mitchell2001157} and references therein).

  Since we are interested in applications of our results for for the analysis of  TRIV effects which are proportional to vector polarization of the target ( $\vec{\sigma}\cdot [\vec{k}\times \vec{I}])$ correlation),
we choose the initial geometry where vector polarization of the target  has the simplest form: neutron spin $\vec{\sigma}$ is parallel to the  axis $x$, target spin $\vec{I}$ is parallel to the axis $z$ (the quantization axis), and neutron momentum $\vec{k}$ is going in the direction of the axis $y$. Therefore,
 for our choice of the coordinate system the angle's values in eq.(\ref{tensAmpNeutr}) are $\theta =\pi /2$ and $\phi =\pi/2$.

\section{Analysis of the scattering with  the target  spin $I=7/2$.}

To understand a  structure of the the amplitude (\ref{gamp}) let us consider an example for a scattering on  the target  with spin $I=7/2$.
In  general,  to completely describe  polarization of nucleus with a spin $I$ we need a set of spherical tensors up to the rank of $q=2I$, which results in $q=7$ for $I=7/2$. However,  for low energy neutron scattering the tensor structure of the amplitude is much simpler. This is because only $s$- and $p$-wave resonances are important, and as a consequence,  one cannot have  tensor terms in the amplitude builded from momentum vector with a rank higher then two (for discussion  of the possible contributions from $d$-wave resonances, see appendix \ref{dwave}).
This results in a constraint that the rank of the target spin tensor has a maximum value $q=3$.

Evaluation of eq.(\ref{tensAmpNeutr}) for $q=0$,  $q=1$, $q=2$, and  $q=3$ with  target spin $7/2$ (see appendix \ref{Fappendix}) leads to results that can be summarized in terms of a linear combination of the tensors already listed in eq.(\ref{gamp}) as
\begin{equation}\label{gamp72}
\begin{split}
  f_{7/2} &=P_0 \Bigl( A^{\prime}+C^{\prime}(\vec{\sigma}\cdot \vec{k}) \Bigr) +P_1 \Bigl(
 B^{\prime}(\vec{\sigma}\cdot \vec{I}) +D^{\prime}(\vec{\sigma}\cdot [\vec{k}\times \vec{I}]) + H^{\prime}(\vec{k}\cdot \vec{I}) + K^{\prime}(\vec{\sigma}\cdot \vec{k})(\vec{k}\cdot \vec{I}) \Bigr) \\
    &+ P_2 \Biggl( E^{\prime}\left( (\vec{k}\cdot \vec{I})(\vec{k}\cdot \vec{I})-\frac{1}{3}(\vec{k}\cdot \vec{k})(\vec{I}\cdot \vec{I})\right)
+  F^{\prime}\left( (\vec{\sigma}\cdot \vec{I})(\vec{k}\cdot \vec{I})-\frac{1}{3}(\vec{\sigma}\cdot \vec{k})(\vec{I}\cdot \vec{I})\right)\\
   &+ G^{\prime}(\vec{\sigma}\cdot [\vec{k}\times \vec{I}])(\vec{k}\cdot \vec{I}) \Biggr) \\
   &+ P_3 \Biggl(B_3^{\prime}\Bigl(  (\vec{\sigma} \cdot \vec{I})[(\vec{k} \cdot \vec{I})(\vec{k} \cdot \vec{I})-\frac{1}{3}(\vec{k} \cdot \vec{k})(\vec{I} \cdot \vec{I})]
        +\frac{2}{5}(\vec{k} \cdot \vec{I})[(\vec{\sigma} \cdot \vec{I})(\vec{k} \cdot \vec{I})\frac{1}{3}+(\vec{\sigma} \cdot \vec{k})(\vec{I} \cdot \vec{I})] \\
        &-\frac{4}{5}(\vec{k} \cdot \vec{I})[(\vec{\sigma} \cdot \vec{I})(\vec{k} \cdot \vec{I})-\frac{1}{3}(\vec{\sigma} \cdot \vec{k})(\vec{I} \cdot \vec{I})]\Bigr) \Biggr) ,
   \end{split}
\end{equation}
where the primed coefficients are defined by the following expressions:
\begin{equation}\label{Apr}
\begin{split}
A^{\prime}= & \frac{i}{32 k  }  \Bigl( 7\left\langle 3, 0 \middle\vert R^{3}\middle\vert 3,0 \right\rangle +9\left\langle 4, 0 \middle\vert R^{4}\middle\vert 4,0 \right\rangle +7\left\langle 3, 1 \middle\vert R^{3}\middle\vert 3,1 \right\rangle \\
&+9\left\langle 3, 1 \middle\vert R^{4}\middle\vert 3,1 \right\rangle
+7\left\langle 4, 1 \middle\vert R^{3}\middle\vert 4,1 \right\rangle +9\left\langle 4, 1 \middle\vert R^{4}\middle\vert 4,1 \right\rangle \Bigr).
\end{split}
\end{equation}

\begin{equation}\label{Cpr}
\begin{split}
C^{\prime}=& \frac{i}{64 k  }
\Bigl( 7(\left\langle 3, 0 \middle\vert R^{3}\middle\vert 3,1 \right\rangle +\left\langle 3, 1 \middle\vert R^{3}\middle\vert 3,0 \right\rangle  )
 -7 \sqrt{3}(\left\langle 3, 0 \middle\vert R^{3}\middle\vert 4,1 \right\rangle +\left\langle 4, 1 \middle\vert R^{3}\middle\vert 3,0 \right\rangle ) \\
 &+3 \sqrt{21}(\left\langle 4, 0 \middle\vert R^{4}\middle\vert 3,1 \right\rangle +\left\langle 3, 1 \middle\vert R^{4}\middle\vert 4,0 \right\rangle   ) -3 \sqrt{15}(\left\langle 4, 0 \middle\vert R^{4}\middle\vert 4,1 \right\rangle
  +\left\langle 4, 1 \middle\vert R^{4}\middle\vert 4,0 \right\rangle ) \Bigr).
\end{split}
\end{equation}

\begin{equation}\label{Bpr}
\begin{split}
B^{\prime}=& -\frac{i}{32 k  }  \Biggl( 7\left\langle 3, 0 \middle\vert R^{3}\middle\vert 3,0 \right\rangle - \left\langle 4, 0 \middle\vert R^{4}\middle\vert 4,0 \right\rangle \\
 &+\frac{21}{4} \left\langle 3, 1 \middle\vert R^{3}\middle\vert 3,1 \right\rangle +\frac{21}{4\sqrt{3}} (\left\langle 3, 1 \middle\vert R^{3}\middle\vert 4,1 \right\rangle +\left\langle 4, 1 \middle\vert R^{3}\middle\vert 3,1 \right\rangle )  \\
 &-\frac{9}{20} \sqrt{35}(\left\langle 3, 1 \middle\vert R^{4}\middle\vert 4,1 \right\rangle + \left\langle 4, 1 \middle\vert R^{4}\middle\vert 3,1 \right\rangle  )  -\frac{91}{12} \left\langle 4, 1 \middle\vert R^{3}\middle\vert 4,1 \right\rangle \\
 & +\frac{39}{4} \left\langle 3, 1 \middle\vert R^{4}\middle\vert 3,1 \right\rangle-\frac{63}{20} \left\langle 4, 1 \middle\vert R^{4}\middle\vert 4,1 \right\rangle  \Biggr).
\end{split}
\end{equation}

\begin{equation}\label{Dpr}
\begin{split}
D^{\prime}=& \frac{1}{16 k  }\Biggl( \frac{7}{\sqrt{3}}(\left\langle 3, 0 \middle\vert R^{3}\middle\vert 4,1 \right\rangle -\left\langle 4, 1 \middle\vert R^{3}\middle\vert 3,0 \right\rangle  ) +\sqrt{21}(\left\langle 4, 0 \middle\vert R^{4}\middle\vert 3,1 \right\rangle -\left\langle 3, 1 \middle\vert R^{4}\middle\vert 4,0 \right\rangle ) \Biggr).
\end{split}
\end{equation}

\begin{equation}\label{Hpr}
\begin{split}
H^{\prime}=& -\frac{i }{64 k  } \Biggl( 21(\left\langle 3, 0 \middle\vert R^{3}\middle\vert 3,1 \right\rangle + \left\langle 3, 1 \middle\vert R^{3}\middle\vert 3,0 \right\rangle ) -\frac{7}{\sqrt{3}}(\left\langle 3, 0 \middle\vert R^{3}\middle\vert 4,1 \right\rangle  +\left\langle 4, 1 \middle\vert R^{3}\middle\vert 3,0 \right\rangle  ) \\
  &+\sqrt{21}(\left\langle 4, 0 \middle\vert R^{4}\middle\vert 3,1 \right\rangle +\left\langle 3, 1 \middle\vert R^{4}\middle\vert 4,0 \right\rangle ) +7 \sqrt{15}(\left\langle 4, 0 \middle\vert R^{4}\middle\vert 4,1 \right\rangle
    +\left\langle 4, 1 \middle\vert R^{4}\middle\vert 4,0 \right\rangle ) \Biggr) .
\end{split}
\end{equation}

\begin{equation}\label{Kpr}
\begin{split}
K^{\prime}=& -\frac{3i }{128 k  } \Biggl( 7\left\langle 3, 1 \middle\vert R^{3}\middle\vert 3,1 \right\rangle -7 \sqrt{3}(\left\langle 3, 1 \middle\vert R^{3}\middle\vert 4,1 \right\rangle +\left\langle 4, 1 \middle\vert R^{3}\middle\vert 3,1 \right\rangle  )  \\
   & +\frac{77}{9\sqrt{3}} \left\langle 4, 1 \middle\vert R^{3}\middle\vert 4,1 \right\rangle-3\left\langle 3, 1 \middle\vert R^{4}\middle\vert 3,1 \right\rangle +\frac{9}{5} \sqrt{35}(\left\langle 3, 1 \middle\vert R^{4}\middle\vert 4,1 \right\rangle + \left\langle 4, 1 \middle\vert R^{4}\middle\vert 3,1 \right\rangle )
    \\
    &  -\frac{77}{5} \left\langle 4, 1 \middle\vert R^{4}\middle\vert 4,1 \right\rangle \Biggr) .
\end{split}
\end{equation}

\begin{equation}\label{Epr}
\begin{split}
E^{\prime}= & \frac{i 9 }{256 k  } \Biggl( \frac{35}{3}   \left\langle 3, 1 \middle\vert R^{3}\middle\vert 3,1 \right\rangle    - \frac{7}{\sqrt{3}} ( \left\langle 3, 1 \middle\vert R^{3}\middle\vert 4,1 \right\rangle +\left\langle 4, 1 \middle\vert R^{3}\middle\vert 3,1 \right\rangle  ) \\
   & -\frac{77}{9} \left\langle 4, 1 \middle\vert R^{3}\middle\vert 4,1 \right\rangle +\frac{3}{5} \sqrt{35}( \left\langle 3, 1 \middle\vert R^{4}\middle\vert 4,1 \right\rangle  + \left\langle 4, 1 \middle\vert R^{4}\middle\vert 3,1 \right\rangle )\\
   & -5\left\langle 3, 1 \middle\vert R^{4}\middle\vert 3,1 \right\rangle
   +\frac{77}{5} \left\langle 4, 1 \middle\vert R^{4}\middle\vert 4,1 \right\rangle \Bigr).
\end{split}
\end{equation}

\begin{equation}\label{Fpr}
\begin{split}
F^{\prime}=&\frac{3i}{320 k  }\Biggl(35(\left\langle 3, 0 \middle\vert R^{3}\middle\vert 3,1 \right\rangle +\left\langle 3, 1 \middle\vert R^{3}\middle\vert 3,0 \right\rangle  ) +\frac{35}{\sqrt{3}}(\left\langle 3, 0 \middle\vert R^{3}\middle\vert 4,1 \right\rangle +\left\langle 4, 1 \middle\vert R^{3}\middle\vert 3,0 \right\rangle ) \\
 &-5\sqrt{21}(\left\langle 4, 0 \middle\vert R^{4}\middle\vert 3,1 \right\rangle +\left\langle 3, 1 \middle\vert R^{4}\middle\vert 4,0 \right\rangle ) -7 \sqrt{15}(\left\langle 4, 0 \middle\vert R^{4}\middle\vert 4,1 \right\rangle
    +\left\langle 4, 1 \middle\vert R^{4}\middle\vert 4,0 \right\rangle ) \Biggr).
\end{split}
\end{equation}

\begin{equation}\label{Gpr}
\begin{split}
G^{\prime}=&-\frac{3 }{32 \sqrt{5} k \pi } \Biggl(7 \sqrt{\frac{5}{3}} (\left\langle 3, 1 \middle\vert R^{3}\middle\vert 4,1 \right\rangle -\left\langle 4, 1 \middle\vert R^{3}\middle\vert 3,1 \right\rangle ) \\
    &-3 \sqrt{7}(\left\langle 3, 1 \middle\vert R^{4}\middle\vert 4,1 \right\rangle -\left\langle 4, 1 \middle\vert R^{4}\middle\vert 3,1 \right\rangle )
    \Biggr) .
\end{split}
\end{equation}

\begin{equation}\label{B3pr}
\begin{split}
B^{\prime}_{3}=& \frac{9i}{256 k  }   \Biggl(  7\left\langle 3, 1 \middle\vert R^{3}\middle\vert 3,1 \right\rangle +(\left\langle 3, 1 \middle\vert R^{3}\middle\vert 4,1 \right\rangle +\frac{7}{\sqrt{3}}\left\langle 4, 1 \middle\vert R^{3}\middle\vert 3,1 \right\rangle  ) \\
& +     \frac{7}{3} \left\langle 4, 1 \middle\vert R^{3}\middle\vert 4,1 \right\rangle
 -3\sqrt{\frac{7}{5}}(\left\langle 3, 1 \middle\vert R^{4}\middle\vert 4,1 \right\rangle +\left\langle 4, 1 \middle\vert R^{4}\middle\vert 3,1 \right\rangle  )  \\
     &-3\left\langle 3, 1 \middle\vert R^{4}\middle\vert 3,1 \right\rangle  -\frac{21}{5} \left\langle 4, 1 \middle\vert R^{4}\middle\vert 4,1 \right\rangle
      \Biggr).
\end{split}
\end{equation}
Since we use a rather particular  spin value $I=7/2$, the above expressions present a parten for a general amplitude structure.

\section{Scattering of polarized neutrons on polarized $^{139}La$ target.}

Let us apply the obtained results  for the case of $^{139}La$ which has a spin $I=7/2$, and
  consider scattering of polarized neutrons on polarized $^{139}La$ target in the vicinity of $p$-resonance $E_p=0.734 eV$. The resonance structure \cite{Mughabghab} of $^{139}La$ shows that there are  two nearest $s$ wave resonances with $E_{s0}=-48.63 eV$  and $E_{s1}=72.3eV$, and with spins $J=4$ and $J=3$, correspondingly. Since $p$-wave resonance has spin $J=4$, it can be mixed only with the negative energy $s0$ resonance. Therefore, we can neglect symmetry violating amplitudes with a total spin $J=3$. Thus,   there are only four parity violating amplitudes  with a total spin $J=4$
\begin{eqnarray}
 \left\langle 4 0 \middle\vert R^{4} \middle\vert 3 1 \right\rangle
&=&
 \frac
{x_S\sqrt{\Gamma^{\rm n}_{s0}}(-iv+w)\sqrt{\Gamma^{\rm n}_{p}}}
{(E-E_p+i\Gamma_p/2)(E-E_{s0}+i\Gamma_{s0}/2)}
e^{i\delta_{s0}} \nonumber \\
 \left\langle 4 0 \middle\vert R^{4} \middle\vert 4 1 \right\rangle
&=&
 \frac
{y_S\sqrt{\Gamma^{\rm n}_{s0}}(-iv+w)\sqrt{\Gamma^{\rm n}_{p}}}
{(E-E_p+i\Gamma_p/2)(E-E_{s0}+i\Gamma_{s0}/2)}
e^{i\delta_{s0}} \nonumber \\
 \left\langle 3 1 \middle\vert R^{4} \middle\vert 40 \right\rangle
&=&
 \frac
{x_S\sqrt{\Gamma^{\rm n}_{s0}}(-iv-w)\sqrt{\Gamma^{\rm n}_{p}}}
{(E-E_p+i\Gamma_p/2)(E-E_{s0}+i\Gamma_{s0}/2)}
e^{-i\delta_{s0}} \nonumber \\
 \left\langle 4 1 \middle\vert R^{4} \middle\vert 4 0 \right\rangle
&=&
 \frac
{y_S\sqrt{\Gamma^{\rm n}_{s0}}(-iv-w)\sqrt{\Gamma^{\rm n}_{p}}}
{(E-E_p+i\Gamma_p/2)(E-E_{s0}+i\Gamma_{s0}/2)}
e^{-i\delta_{s0}} .
\end{eqnarray}
Thre are  six parity conserving amplitudes
\begin{eqnarray}
  \left\langle 4 0 \middle\vert R^{4} \middle\vert 4 0 \right\rangle
&=&
i \frac
{\Gamma^{\rm n}_{s0}}
{E-E_{s0}+i\Gamma_{s0}/2}
e^{2i\delta_{s0}}
-2ie^{i\delta_{s0}}\sin\delta_{s0} \nonumber \\
  \left\langle 3 0 \middle\vert R^{3} \middle\vert 3 0 \right\rangle
&=&
i \frac
{\Gamma^{\rm n}_{s1}}
{E-E_{s1}+i\Gamma_{s1}/2}
e^{2i\delta_{s1}}
-2ie^{i\delta_{s1}}\sin\delta_{s1} \nonumber  \\
  \left\langle 3 1 \middle\vert R^{4} \middle\vert 3 1 \right\rangle
&=&
i \frac
{x^2_S\Gamma^{\rm n}_{p}}
{E-E_{p}+i\Gamma_{p}/2}
 \nonumber  \\
\left\langle 4 1 \middle\vert R^{4} \middle\vert 4 1 \right\rangle
&=&
i \frac
{y^2_S\Gamma^{\rm n}_{p}}
{E-E_{p}+i\Gamma_{p}/2}
 \nonumber   \\
\left\langle 3 1 \middle\vert R^{4} \middle\vert 4 1 \right\rangle
&=&
i \frac
{x_S y_S\Gamma^{\rm n}_{p}}
{E-E_{p}+i\Gamma_{p}/2}
 \nonumber   \\
 \left\langle 4 1 \middle\vert R^{4} \middle\vert 3 1 \right\rangle
&=&
i \frac
{x_S y_S\Gamma^{\rm n}_{p}}
{E-E_{p}+i\Gamma_{p}/2}
 \end{eqnarray}
that make main contributions in the vicinity of $p$-resonance. Also, for slow neutrons we can neglect exponentials with phases in the above expressions.

For the case of phenomenological TRIV and parity conserving   interactions  (TVPC), corresponding to the term  $G^{\prime}$,  there are four possible amplitudes for $^{139}La$ target which contribute to eq.(\ref{Gpr}) by two matrix element differences $(\left\langle 3, 2 \middle\vert R^{4}_T\middle\vert 4,0 \right\rangle -\left\langle 4, 0 \middle\vert R^{4}_T\middle\vert 3,2 \right\rangle )$ and $(\left\langle 3, 1 \middle\vert R^{4}_T\middle\vert 4,1 \right\rangle - \left\langle 4, 1 \middle\vert R^{4}_T\middle\vert 3,1 \right\rangle )$. Assuming only  contributions from a compound resonance mixing (see detailed discussions in \cite{Bunakov:1988eb,Gudkov:1991qc,Barabanov:2005tc} ), one can write these differences as:
\begin{eqnarray}
(\left\langle 3, 1 \middle\vert R^{4}_T\middle\vert 4,1 \right\rangle - \left\langle 4, 1 \middle\vert R^{4}_T\middle\vert 3,1 \right\rangle)
&=&
 \frac
{iv^{pp}_T((\Gamma(3)^{\rm n}_{p1}\Gamma(4)^{\rm n}_{p2})^{1/2} - (\Gamma(4)^{\rm n}_{p1}\Gamma(3)^{\rm n}_{p2})^{1/2} )}
{(E-E_{p1}+i\Gamma_{p1}/2)(E-E_{p2}+i\Gamma_{p2}/2)}
 \nonumber \\
(\left\langle 3, 2 \middle\vert R^{4}_T\middle\vert 4,0 \right\rangle -\left\langle 4, 0 \middle\vert R^{4}_T\middle\vert 3,2 \right\rangle )
&=&
\frac
{iv^{sd}_T((\Gamma(3)^{\rm n}_{d}\Gamma(4)^{\rm n}_{s})^{1/2} - (\Gamma(4)^{\rm n}_{d}\Gamma(3)^{\rm n}_{s})^{1/2} )}
{(E-E_{d}+i\Gamma_{d}/2)(E-E_{s}+i\Gamma_{s}/2)} e^{i\delta_{s}},
\end{eqnarray}
where $v^{pp}_T$ and  $v^{sd}_T$ phenomenological TVPC matrix elements \cite{Gudkov:1991qc}, and $\Gamma(S)^{\rm n}_{k}$ are partial neutron decay width for $k$'s resonance corresponding to spin channel $S$. We can see from these expressions that for the existence of already very small coefficient $G^{\prime}$ (due to a naturally small value of phenomenological TVPC interactions \cite{TVPC_R-M}) one has to have either additional $p$-wave resonance, or $d$-wave resonance. Since we are interested in the region in the vicinity of $p$-wave resonance, the possible contributions from $s$-$d$ mixture can be neglected, therefore only first difference of matrix elements in the above expression  can be taken into account.  For the completeness of the description of the neutron propagation through the polarized target we provide the expression for $G^{\prime}$ coefficient  however we  neglect TVPC correlations in the further analysis. (For experimental constrain of this term from neutron scattering on aligned holmium see \cite{Huffman:1996zz,Huffman:1996ix}.)

 Then, the coefficients in the amplitude (\ref{gamp72}) for $^{139}La$ are:
\begin{equation}\label{aqLa}
\begin{split}
A^{\prime}_{La}=&\frac{-1}{32 k  } \Biggl( 7 \frac
{\Gamma^{\rm n}_{s1}}
{E-E_{s1}+i\Gamma_{s1}/2}
 + 9  \frac
{\Gamma^{\rm n}_{s0}}
{E-E_{s0}+i\Gamma_{s0}/2}
 \\
    +&9  \frac
{\Gamma^{\rm n}_{p}}
{E-E_{p}+i\Gamma_{p}/2}   \Biggr)
+ \frac{1}{ 16  }\biggl(9a_{s0} +7a_{s1} \biggr),
\end{split}
\end{equation}

\begin{equation}\label{bqLa}
\begin{split}
B^{\prime}_{La}=&\frac{-1}{32 k  } \Biggl( 7 \frac
{\Gamma^{\rm n}_{s0}}
{E-E_{s0}+i\Gamma_{s0}/2} -7 \frac
{\Gamma^{\rm n}_{s1}}
{E-E_{s1}+i\Gamma_{s1}/2} \\
    +&  \frac
{\Gamma^{\rm n}_{p}}
{E-E_{p}+i\Gamma_{p}/2} \biggl( -\frac{39}{4}x^2_S +   \frac{9}{2}\sqrt{\frac{7}{5}}x_S y_S +\frac{63}{20}y^2_S  \biggr)    \Biggr) \\
+& \frac{7}{ 16 }\biggl(a_{s0} - a_{s1} \biggr),
\end{split}
\end{equation}

\begin{equation}\label{cqLa}
\begin{split}
C^{\prime}_{La}=\frac{3 \sqrt{3}}{32 k  }
  \frac
{\sqrt{\Gamma^{\rm n}_{s0}}v\sqrt{\Gamma^{\rm n}_{p}}}
{(E-E_p+i\Gamma_p/2)(E-E_{s0}+i\Gamma_{s0}/2)} \biggl( \sqrt{7}x_S  -\sqrt{5} y_S \biggr)
,
\end{split}
\end{equation}

\begin{equation}\label{dqLa}
\begin{split}
D^{\prime}_{La}=&\frac{ \sqrt{21}}{8  k  }
 \frac
{x_S\sqrt{\Gamma^{\rm n}_{s0}}w\sqrt{\Gamma^{\rm n}_{p}}}
{(E-E_p+i\Gamma_p/2)(E-E_{s0}+i\Gamma_{s0}/2)}  ,
\end{split}
\end{equation}

\begin{equation}\label{hqLa}
\begin{split}
H^{\prime}_{La}=-\frac{3 \sqrt{7}}{32 k  }
  \frac
{\sqrt{\Gamma^{\rm n}_{s0}}v\sqrt{\Gamma^{\rm n}_{p}}}
{(E-E_p+i\Gamma_p/2)(E-E_{s0}+i\Gamma_{s0}/2)} \biggl( x_S  +\sqrt{35} y_S \biggr)
,
\end{split}
\end{equation}

\begin{equation}\label{KqLa}
\begin{split}
K^{\prime}_{La}=&-\frac{\sqrt{21}}{128 k }
 \frac
{\Gamma^{\rm n}_{p}}
{E-E_{p}+i\Gamma_{p}/2} \biggl(3 \sqrt{\frac{3}{7}}x^2_S -18 \sqrt{\frac{3}{5}}x_S y_S+
\frac{11}{5} \sqrt{21}y^2_S \biggr),
\end{split}
\end{equation}

\begin{equation}\label{EqLa}
\begin{split}
E^{\prime}_{La}=&\frac{3\sqrt{21}}{256 k  }
 \frac
{\Gamma^{\rm n}_{p}}
{E-E_{p}+i\Gamma_{p}/2} \biggl(5 \sqrt{\frac{3}{7}}x^2_S -6 \sqrt{\frac{3}{5}}x_S y_S-
\frac{11}{5} \sqrt{21}y^2_S \biggr),
\end{split}
\end{equation}

\begin{equation}\label{FqLa}
\begin{split}
F^{\prime}_{La}=&-\frac{3\sqrt{21}}{32 k  } \frac
{\sqrt{\Gamma^{\rm n}_{s0}}v\sqrt{\Gamma^{\rm n}_{p}}}
{(E-E_p+i\Gamma_p/2)(E-E_{s0}+i\Gamma_{s0}/2)} \Bigl( x_S + \sqrt{\frac{7}{5}} y_S \Bigr) ,
\end{split}
\end{equation}

\begin{equation}\label{GqLa}
\begin{split}
G^{\prime}_{La}=\frac{9 \sqrt{7}i}{32 \sqrt{5} k  }
  \frac
{\sqrt{\Gamma^{\rm n}_{p1}}v^{pp}_T\sqrt{\Gamma^{\rm n}_{p}}}
{(E-E_p+i\Gamma_p/2)(E-E_{p1}+i\Gamma_{p1}/2)} \biggl( x_S y_{S1} - x_{S1} y_S \biggr) ,
\end{split}
\end{equation}

\begin{equation}\label{b3qLa}
\begin{split}
(B^{\prime}_3)_{La}=&\frac{189}{256 k  }
 \frac
{\Gamma^{\rm n}_{p}}
{E-E_{p}+i\Gamma_{p}/2} \biggl( \frac{1}{7}x^2_S +\frac{2}{\sqrt{35}}x_S y_S+
\frac{1}{5} y^2_S \biggr),
\end{split}
\end{equation}
where $a_{s0}$, and $a_{s1}$ are neutron scattering lengths with the total spins $J=4$ and $J=3$, correspondingly.

We can see that not all these correlations are equally important for the analysis of neutron spin propagation through the polarized target for experiments to measure the $D^{\prime}$ correlation.  Correlations $A^{\prime}$ and $B^{\prime}$ are related to strong spin independent and spin dependent backgrounds, correspondingly, and $C^{\prime}$ to a weak spin dependent background. However, in spite of the fact that values of  $H^{\prime}$ and $K^{\prime}$ coefficients are about the same order of magnitude  as $C^{\prime}$, they are proportional to $(\vec{k} \cdot \vec{I})$ term. Therefore, they can be well controlled by a precise alignment of the target spin to be perpendicular to the direction of the neutron beam. The coefficients $E^{\prime}$ and $F^{\prime}$  are also of the order of magnitude of $C^{\prime}$, however they are proportional to the second order of the tensor polarization $P_2$, and, therefore can be minimized by creating a pure vector polarization. Term $G^{\prime}$ is very small as it was discussed above, and it is also proportional to the tensor polarization $P_2$. Moreover, $G^{\prime}$ is  proportional to $(\vec{k} \cdot \vec{I})$ product, which gives an additional way to suppressed it by an alignment of the target spin. Finally, $B^{\prime}_3$ being of the same order as  $E^{\prime}$, is proportional to the third order of the tensor polarization $P_3$, which usually is  small.

\section{Neutron spin rotation}

Using expressions for the scattering amplitude $f$ (see, for example eqs. (\ref{gamp}), (\ref{F0vec}), (\ref{F1vec}), (\ref{F2vec}), (\ref{F3vec}), and (\ref{gamp72}) )
we can describe the transmission of polarized neutrons through polarized medium and in external magnetic field $\vec{B}$ by Schr\"{o}dinger's equation (see, for example \cite{Stodolsky:1985ic,Kabir:1988ma,Lamoreaux:1994nd} and references therein) with the effective Hamiltonian (Fermi potential):
\begin{equation}\label{Ham}
H=-\frac{2\pi \hbar ^2}{m_n}Nf-\frac{\mu}{2}(\vec{\sigma}\cdot \vec{B})
\end{equation}
where $m_n$  is the neutron mass, $N$  is the number of scattering centers per unit volume, and $\mu$ is neutron magnetic moment.
Then the evolution operator which relates the initial neutron spinor to the spinor at the distance $y$ is
\begin{equation}\label{evolution}
 U=e^{-i\frac{Hy}{v\hbar}},
\end{equation}
where $v$ is neutron velocity.
Following the approach of Stodolsky \cite{Stodolsky:1985ic} by representing the amplitude  in the form of $2 \times 2$ matrix in neutron-spin space as $f=a +(\vec{\sigma} \cdot \vec{b})$ and using formula $exp(i\vec{\sigma} \cdot \vec{b})=\cos b+i(\vec{\sigma} \cdot \vec{b}) (\sin b)/|b|$, we can calculate coefficients $A(y)$, $B(y)$,  etc, which correspond to $A^{\prime}$, $B^{\prime}$, etc in eqs. (\ref{gamp}) and (\ref{gamp72}) at $y=0$.
For the case of $I>= 3/2$, and, in particular for $I=7/2$, this parametrization leads to
\begin{eqnarray}\label{Stodol}
\nonumber
  a &=& P_0  A^{\prime}+P_1  H^{\prime}(\vec{k}\cdot \vec{I})+P_2  E^{\prime}\left( (\vec{k}\cdot \vec{I})(\vec{k}\cdot \vec{I})-\frac{1}{3}\right) -\frac{P_3B_3^{\prime} }{3}, \\
  \nonumber
  b_i &=& P_0 C^{\prime}k_i  -\mu_{eff} (\vec{B})_i /2
  +P_1 \left(
 B^{\prime}I_i +D^{\prime} [\vec{k}\times \vec{I}]_i  + K^{\prime}k_i(\vec{k}\cdot \vec{I})\right) \\
   &+&  P_2 \left(
 F^{\prime}\left( I_i(\vec{k}\cdot \vec{I})-\frac{1}{3}k_i \right)
   + G^{\prime} [\vec{k}\times \vec{I}]_i(\vec{k}\cdot \vec{I}) \right) \\
   \nonumber
   &+& P_3  B_3^{\prime}\left(  I_i[(\vec{k} \cdot \vec{I})(\vec{k} \cdot \vec{I})]
        +\frac{2}{5}(\vec{k} \cdot \vec{I})[I_i(\vec{k} \cdot \vec{I})\frac{1}{3}+k_i]
        -\frac{4}{5}(\vec{k} \cdot \vec{I})[I_i(\vec{k} \cdot \vec{I})-\frac{1}{3}k_i]\right) ,
\end{eqnarray}
where $\mu_{eff}=\mu m_n/ (4\pi \hbar^2 N)$. (For spin  $I=1$ we should assign $P_3=0$  in the above expression, and for the case of   $I=1/2$  we have both   $P_2=0$ and $P_3=0$.)

One can see that for the perfect alignment of  the magnetic field  along axis $z$ and neutron momentum along axis $y$ the eq. (\ref{Stodol}) transforms to
\begin{eqnarray}\label{Stodolgeo}
\nonumber
  a &=& P_0  A^{\prime} -\frac{P_2  E^{\prime}+P_3B_3^{\prime} }{3}, \\
  \nonumber
  b_1 &=& P_1
 D^{\prime} [\vec{k}\times \vec{I}]_1  , \\
      b_2 &=& P_0 C^{\prime} -
  \frac{ P_2F^{\prime}}{3} \\
  \nonumber
    b_3 &=&   -\mu_{eff} (\vec{B})_3 /2   +P_1  B^{\prime} .
 \end{eqnarray}
The parameter $a$ defers from its value in \cite{Stodolsky:1985ic} by the second term, which can be ignored since  it  is suppressed by a factor $(kR)^2$ and numerically by  small values of the tensor polarizations. The parameters $b_1$ and $b_2$ are exactly the same as for a ``simple'' spin amplitude in \cite{Stodolsky:1985ic}. Therefore, the only important correction to \cite{Stodolsky:1985ic,Lamoreaux:1994nd} analysis of spin rotation in the target with $I>1/2$ is the second term of $b_2$ (for  the target with $I=1/2$ only vector polarization exist, and $P_2=P_3=0$).

Thus, for analysis of neutron spin rotation   we can use the results of \cite{Stodolsky:1985ic} by changing the parameter $b_2$ according to eq.(\ref{Stodolgeo}), and   by changing the coordinate system in \cite{Stodolsky:1985ic,Kabir:1988ma} as $x \rightarrow z$,  $y \rightarrow x$, and  $z \rightarrow y$. Alternatively, one can use the explicit expression for the evolution operator (\ref{evolution}) as a function of a distance $y$, which can be written as
\begin{equation}\label{Uspin}
  U(y)=\exp (i\alpha y)\left[ \cos (\beta y) +i\frac{\vec{(\sigma}\cdot \vec{b})}{b}\sin (\beta y)  \right],
\end{equation}
where  $\alpha = \gamma a$, $\beta_i = \gamma b_i$, $\gamma=2\pi \hbar N/(m_n v)$, and       $\beta =|\vec{\beta}|$.

It should be noted, that the parameter $a$ contributes to a general attenuation of the neutron beam, but value of $b$ is washing out neutron spin component of the amplitude. Therefore, the smaller value means better sensitivity for  spin related observables (TRIV effect, in particular).   The largest part of $b$ come from $b_3$ component, which can be reduced by adjusting external magnetic field (see \cite{Gudkov:2017sye} and references therein).  The second large part comes from $b_2$, which also can be reduced by adjusting the value of the second rank of the target polarization $P_2$.

\section{Conclusions}

We have developed a general systematic approach for the description of low energy neutron spin propagation in the target with arbitrary polarization using irrecusable spherical tensor representation for target polarizations.
 Applying this technique for the case of slow neutrons,  when only $s$- and $p$- wave resonances are important, we demonstrated that for any value of the target spin  only terms up to the third rank of tensor polarization are present in the scattering amplitude, in compare to $2I$-th rank of tensor polarization in a general case. This is because the low power of momenta (up to second rank tensor in our case) cannot be coupled to a higher rank of  spin tensors. Therefore, even for targets with a large value of spin, the number of irreducible terms in the scattering amplitude is less or equal to ten. For a target with spin $I=1/2$ only six irreducible terms exist.

The analysis of neuron spin propagation shows that only seven terms from the possible ten ones are numerically important for the description of the neutron spin propagation, and this number can be even decreased  to four for the target with $I=1/2$.

The obtained results provide the recipe how to extend the existent ``conventional'' approach for the description of neutron spin dynamics in vector polarized target on the case of arbitrary polarized target.

Another important observation is related to the fact that the second rank tensor polarization of the target can be used for the cancelation of neutron spin rotation due to weak interaction. This, combined with the possible cancellation of the strong spin-spin interaction by an external magnetic field, gives the opportunity for  essential increasing of the sensitivity in the search for TRIV.

Finally, using obtained formalism we presented the detailed description of neutron spin propagation in arbitrary polarized $^{139}La$, which is one of the candidates for the target for the future TRIV experiments.

\begin{acknowledgments}
This material is based upon work supported by the U.S. Department of Energy Office of Science, Office of Nuclear Physics program under Award No. DE-SC0015882.
\end{acknowledgments}

\appendix

\section{Spin Polarization}
\label{tensor}

\noindent
To specify polarization in the spherical representation we use the statistical tensors $\widetilde{\tau}_{\kappa q}$ which are defined as  the expectation values of irreducible tensor spin operators
\begin{equation}\label{statTens}
\widetilde{\tau}_{\kappa q}=\langle \tau_{\kappa q} \rangle
\end{equation}

The spin operator corresponding to the spin $j$ is defined \cite{Lakin} by
\begin{equation}
(\tau^j_{kq})_{m'm} = \braoket{jm'}{\tau^j_{kq}}{jm}
	= \sqrt{2k+1} \cgc{j}{m}{k}{q}{j}{m'}
.
\end{equation}

It should be noted that tensors $\widetilde{\tau}_{\kappa q}$ have different normalization in compare to tensors $t_{\kappa q}$ in eq.(\ref{denMart}),
which are defined as \cite{Varshalovich}
\begin{equation}
(t_{kq})_{m'm} =  \sqrt{\frac{2k+1}{2j+1}} \cgc{j}{m}{k}{q}{j}{m'}
.
\end{equation}

We define the population of each magnetic substate $m$ of the spin $\bm{I}$ as
\begin{eqnarray}
N_m = \braket{Im}{Im}
,\quad
\sum_{m=-I}^{I} N_m = 1
.
\end{eqnarray}
Then we can calculate
\begin{eqnarray}
\overline{\tau}^I_k = \sum_{q=-I}^{I} \braoket{Iq}{\tau^{I}_{k0}}{Iq},
\end{eqnarray}
for $k$ in the range of $1 \le k \le 2I$.

Since the vector polarization $P^{I}_1$ is commonly defined to be proportional to $\overline{\tau}^{I}_1$, we write
\begin{eqnarray}\label{defpol}
P^{I}_1 = c^I_1 \overline{\tau}^{I}_1,
\end{eqnarray}
where the constant $c^I_1$ is obtained from the requirement that  $P^{I}_1=1$ for $N_I=1$:
\begin{eqnarray}
c^I_1 &=& \frac{1}{(\tau^{I}_{1,0})_{II}}.
\end{eqnarray}
Then the vector polarization is uniquely defined as
\begin{eqnarray}
P^{I}_1 = \frac{\overline{\tau}^{I}_1}{(\tau^{I}_{1,0})_{II}},
\end{eqnarray}
and the general tensor polarizations are
\begin{eqnarray}
P^{I}_{k} = c^I_k \overline{\tau}^{I}_k
\end{eqnarray}
with parameters $c^I_k$ determined from the condition that  $P^{I}_k=1$ for $N_I=1$.
Thus polarization for arbitrary cases can be uniquely defined as
\begin{eqnarray}
P^{I}_k = \frac{\overline{\tau}^{I}_k}{(\tau^{I}_{k,0})_{II}}
\end{eqnarray}
with  normalization coefficients
\begin{eqnarray}\label{polcoeff}
\frac{1}{(\tau^{I}_{k,0})_{II}}
	&=&
	\frac{1}{\sqrt{(2I+1)(2k+1)}}
	\sqrt{\frac
		{(2I-k)!}
		{(2I)!}
	}
	\sqrt{\frac
		{(2I+k+1)!}
		{(2I)!}
	}.
\end{eqnarray}

This leads to  explicit expressions for $P^I_k$ as:
\begin{eqnarray}
P^{\frac12}_1 &=& N_{\frac12} - N_{-\frac12} \nonumber
\end{eqnarray}
\begin{eqnarray}
P^{1}_1 &=& N_{1} - N_{-1} \nonumber\\
P^{1}_2 &=& N_{1} - 2N_{0} + N_{-1} \nonumber
\end{eqnarray}
\begin{eqnarray}
P^{\frac32}_1 &=& N_{\frac32} + \frac13 N_{\frac12} - \frac13 N_{-\frac12} - N_{-\frac32} \nonumber\\
P^{\frac32}_2 &=& N_{\frac32} - N_{\frac12} - N_{-\frac12} + N_{-\frac32} \nonumber\\
P^{\frac32}_3 &=& N_{\frac32} - 3N_{\frac12} + 3N_{-\frac12} - N_{-\frac32} \nonumber
\end{eqnarray}
\begin{eqnarray}
P^{2}_1 &=& N_{2} +\frac14 N_{1} -\frac14 N_{-1} - N_{-2} \nonumber\\
P^{2}_2 &=& N_{2} -\frac12 N_{1} - N_{0} -\frac12 N_{-1} + N_{-1} \nonumber\\
P^{2}_3 &=& N_{2} -2 N_{1} +2 N_{-1} - N_{-1} \nonumber\\
P^{2}_4 &=& N_{2} -4 N_{1} +6 N_{0} -4 N_{-1} + N_{-1} \nonumber
\end{eqnarray}
\begin{eqnarray}
P^{\frac52}_1 &=& N_{\frac52} +\frac35 N_{\frac32} +\frac15 N_{\frac12} -\frac15 N_{-\frac12} -\frac35 N_{-\frac32} -N_{-\frac52}
	\nonumber\\
P^{\frac52}_2 &=& N_{\frac52} -\frac15 N_{\frac32} -\frac45 N_{\frac12} -\frac45 N_{-\frac12} -\frac15 N_{-\frac32} +N_{-\frac52}
	\nonumber\\
P^{\frac52}_3 &=& N_{\frac52} -\frac75 N_{\frac32} -\frac45 N_{\frac12} +\frac45 N_{-\frac12} +\frac75 N_{-\frac32} -N_{-\frac52}
	\nonumber\\
P^{\frac52}_4 &=& N_{\frac52} -3N_{\frac32} +2N_{\frac12} +2N_{-\frac12} -3N_{-\frac32} +N_{-\frac52}
	\nonumber\\
P^{\frac52}_5 &=& N_{\frac52} -5N_{\frac32} +10N_{\frac12} -10N_{-\frac12} +5N_{-\frac32} -N_{-\frac52}
	\nonumber
\end{eqnarray}
\begin{eqnarray}
P^{3}_1 &=& N_{3} +\frac23 N_{2} +\frac13 N_{1} -\frac13 N_{-1} -\frac23 N_{-2} -N_{-3}
	\nonumber\\
P^{3}_2 &=& N_{3} -\frac35 N_{1} -\frac45 N_{0} -\frac35 N_{-1} +N_{-3}
	\nonumber\\
P^{3}_3 &=& N_{3} -N_{2} -N_{1} +N_{-1} +N_{-2} -N_{-3}
	\nonumber\\
P^{3}_4 &=& N_{3} -\frac73 N_{2} +\frac13 N_{1} +2 N_{0} +\frac13N_{-1} -\frac73 N_{-2} +N_{-3}
	\nonumber\\
P^{3}_5 &=& N_{3} -4N_{2} +5N_{1} -5N_{-1} +4N_{-2} -N_{-3}
	\nonumber\\
P^{3}_6 &=& N_{3} -6N_{2} +15N_{1} -20N_{0} +15N_{-1} -6N_{-2} +N_{-3}
	\nonumber
\end{eqnarray}
\begin{eqnarray}
P^{\frac72}_1 &=& N_{\frac72} +\frac57 N_{\frac52} +\frac37 N_{\frac32} +\frac17 N_{\frac12} -\frac17 N_{-\frac12} -\frac37 N_{-\frac32} -\frac57 N_{-\frac52} -N_{-\frac72}
	\nonumber\\
P^{\frac72}_2 &=& N_{\frac72} +\frac17 N_{\frac52} -\frac37 N_{\frac32} -\frac57 N_{\frac12} -\frac57 N_{-\frac12} -\frac37 N_{-\frac32} +\frac17 N_{-\frac52} +N_{-\frac72}
	\nonumber\\
P^{\frac72}_3 &=& N_{\frac72} -\frac57 N_{\frac52} -N_{\frac32} -\frac37 N_{\frac12} +\frac37 N_{-\frac12} +N_{-\frac32} +\frac57 N_{-\frac52} -N_{-\frac72}
	\nonumber\\
P^{\frac72}_4 &=& N_{\frac72} -\frac{13}{7} N_{\frac52} -\frac37 N_{\frac32} +\frac97 N_{\frac12} +\frac97 N_{-\frac12} -\frac37 N_{-\frac32} -\frac{13}{7} N_{-\frac52} +N_{-\frac72}
	\nonumber\\
P^{\frac72}_5 &=& N_{\frac72} -\frac{23}{7} N_{\frac52} -\frac{17}{7} N_{\frac32} +\frac{15}{7} N_{\frac12} -\frac{15}{7} N_{-\frac12} +\frac{17}{7} N_{-\frac32} +\frac{23}{7} N_{-\frac52} -N_{-\frac72}
	\nonumber\\
P^{\frac72}_6 &=& N_{\frac72} -5N_{\frac52} +9N_{\frac32} -5N_{\frac12} -5N_{-\frac12} +9N_{-\frac32} -5N_{-\frac52} +N_{-\frac72}
	\nonumber\\
P^{\frac72}_7 &=& N_{\frac72} -7N_{\frac52} +21N_{\frac32} -35N_{\frac12} +35N_{-\frac12} -21N_{-\frac32} +7N_{-\frac52} -N_{-\frac72}
	\nonumber
\end{eqnarray}

\section{Some useful expressions with fixed values of $M$.}
\label{Mappendix}

For numerical calculations it may be convenient to have
 expressions for for $A^{\prime}$, $B^{\prime}$, etc coefficients for specific values of the projection of the target spins on the axis $z$. Using eq.(\ref{genAmp}) on can get these expressions for a particular values of the target spin $I$.

Such, for the target spin $I=1/2$ and spin projection $M=\pm 1/2$:
\begin{equation}\label{aM12}
\begin{split}
A^{\prime}_{1/2\pm1/2}=& \frac{i}{8 k  } \biggl( \left\langle 0, 0\middle\vert R^{0}\middle\vert 0,0 \right\rangle+3 \left\langle 1, 0\middle\vert R^{1}\middle\vert 1,0 \right\rangle
+3  \left\langle 0, 1\middle\vert R^{1}\middle\vert 0,1 \right\rangle \\
+& \left\langle 1, 1\middle\vert R^{0}\middle\vert 1,1 \right\rangle+3  \left\langle 1, 1\middle\vert R^{1}\middle\vert 1,1 \right\rangle \biggr)
\end{split}
\end{equation}
For the target spin $I=1$ and spin projection $M=\pm 1$:
\begin{equation}\label{aM11}
\begin{split}
A^{\prime}_{1\pm 1}=&\frac{i}{120 k  } \Biggl( 20\left\langle \frac{1}{2}, 0 \middle\vert R^{\frac{1}{2}}\middle\vert \frac{1}{2},0 \right\rangle +40\left\langle \frac{3}{2}, 0 \middle\vert R^{\frac{3}{2}}\middle\vert \frac{3}{2},0 \right\rangle +20\left\langle \frac{1}{2}, 1 \middle\vert R^{\frac{1}{2}}\middle\vert \frac{1}{2},1 \right\rangle \\
 +& 5 \sqrt{2}\left\langle \frac{1}{2}, 1 \middle\vert R^{\frac{1}{2}}\middle\vert \frac{3}{2},1 \right\rangle +40\left\langle \frac{1}{2}, 1 \middle\vert R^{\frac{3}{2}}\middle\vert \frac{1}{2},1 \right\rangle -2 \sqrt{5}\left\langle \frac{1}{2}, 1 \middle\vert R^{\frac{3}{2}}\middle\vert \frac{3}{2},1 \right\rangle \\
 +& 5 \sqrt{2}\left\langle \frac{3}{2}, 1 \middle\vert R^{\frac{1}{2}}\middle\vert \frac{1}{2},1 \right\rangle +25\left\langle \frac{3}{2}, 1 \middle\vert R^{\frac{1}{2}}\middle\vert \frac{3}{2},1 \right\rangle -2 \sqrt{5}\left\langle \frac{3}{2}, 1 \middle\vert R^{\frac{3}{2}}\middle\vert \frac{1}{2},1 \right\rangle \\
 +& 32\left\langle \frac{3}{2}, 1 \middle\vert R^{\frac{3}{2}}\middle\vert \frac{3}{2},1 \right\rangle \Biggr)
\end{split}
\end{equation}
For the target spin $I=1$ and spin projection $M=0$:
\begin{equation}\label{aM10}
\begin{split}
A^{\prime}_{1 0}=&\frac{i}{120 k } \Biggl( 20\left\langle \frac{1}{2}, 0 \middle\vert R^{\frac{1}{2}}\middle\vert \frac{1}{2},0 \right\rangle +40\left\langle \frac{3}{2}, 0 \middle\vert R^{\frac{3}{2}}\middle\vert \frac{3}{2},0 \right\rangle +20\left\langle \frac{1}{2}, 1 \middle\vert R^{\frac{1}{2}}\middle\vert \frac{1}{2},1 \right\rangle \\
 -& 10 \sqrt{2}\left\langle \frac{1}{2}, 1 \middle\vert R^{\frac{1}{2}}\middle\vert \frac{3}{2},1 \right\rangle +40\left\langle \frac{1}{2}, 1 \middle\vert R^{\frac{3}{2}}\middle\vert \frac{1}{2},1 \right\rangle +4 \sqrt{5}\left\langle \frac{1}{2}, 1 \middle\vert R^{\frac{3}{2}}\middle\vert \frac{3}{2},1 \right\rangle \\
 -& 10 \sqrt{2}\left\langle \frac{3}{2}, 1 \middle\vert R^{\frac{1}{2}}\middle\vert \frac{1}{2},1 \right\rangle +10\left\langle \frac{3}{2}, 1 \middle\vert R^{\frac{1}{2}}\middle\vert \frac{3}{2},1 \right\rangle +4 \sqrt{5}\left\langle \frac{3}{2}, 1 \middle\vert R^{\frac{3}{2}}\middle\vert \frac{1}{2},1 \right\rangle \\
 +& 56\left\langle \frac{3}{2}, 1 \middle\vert R^{\frac{3}{2}}\middle\vert \frac{3}{2},1 \right\rangle \Biggr)
\end{split}
\end{equation}
or the target spin $I>1$ and spin projection $M$:
\begin{equation}\label{aMIany}
\begin{split}
A^{\prime}_{IM}=&\frac{i}{4 (2 I+1)^2 k  } \Biggl( 2I(2I+1) \left\langle (I-\frac{1}{2}), 0\middle\vert R^{I-\frac{1}{2}}\middle\vert (I-\frac{1}{2}),0 \right\rangle \\
+&2(I+1)(2I+1) \left\langle (I+\frac{1}{2}), 0 \middle\vert R^{I+\frac{1}{2}}\middle\vert (I+\frac{1}{2}),0 \right\rangle \\
+&\frac{6  \bigl(2 I^3+2 M^2-I \bigl(1+2 M^2\bigr)\bigr) }{ (2 I-1)  } \left\langle (I-\frac{1}{2}), 1 \middle\vert R^{I-\frac{1}{2}}\middle\vert (I-\frac{1}{2}),1 \right\rangle  \\
-&\frac{3  \bigl(I+I^2-3 M^2\bigr)}{  \sqrt{(I+1) (2 I-1)} } \bigl( \left\langle (I-\frac{1}{2}), 1\middle\vert R^{I-\frac{1}{2}}\middle\vert (I+\frac{1}{2}),1 \right\rangle + \left\langle (I+\frac{1}{2}), 1 \middle\vert R^{I-\frac{1}{2}}\middle\vert (I-\frac{1}{2}),1 \right\rangle  \bigr) \\
+&\frac{3  \bigl(I+2 I^2+I^3-M^2+I M^2\bigr) }{ I  } \left\langle (I-\frac{1}{2}), 1 \middle\vert R^{I+\frac{1}{2}}\middle\vert (I-\frac{1}{2}),1 \right\rangle  \\
+&\frac{3  \bigl(I+I^2-3 M^2\bigr)}{ \sqrt{I (2 I+3)} } \bigl( \left\langle (I-\frac{1}{2}), 1 \middle\vert R^{I+\frac{1}{2}}\middle\vert (I+\frac{1}{2}),1 \right\rangle   + \left\langle (I+\frac{1}{2}), 1 \middle\vert R^{I+\frac{1}{2}}\middle\vert (I-\frac{1}{2}),1 \right\rangle \bigr) \\
+&\frac{3  \bigl(I^2+I^3+2 M^2+I M^2\bigr)}{(I+1)  } \left\langle (I+\frac{1}{2}), 1 \middle\vert R^{I-\frac{1}{2}}\middle\vert (I+\frac{1}{2}),1 \right\rangle  \\
+&\frac{6  \bigl(1+6 I^2+2 I^3-4 M^2+I \bigl(5-2 M^2\bigr)\bigr) }{ (2 I+3)  }  \left\langle (I+\frac{1}{2}), 1 \middle\vert R^{I+\frac{1}{2}}\middle\vert (I+\frac{1}{2}),1 \right\rangle \Biggr)
\end{split}
\end{equation}
For the target spin $I>1$ with $M=I$:
\begin{equation}\label{aMI}
\begin{split}
A^{\prime}_I=&\frac{i}{4 (2 I+1)^2 k } \Biggl( 2 I (2I+1) \left\langle (I-\frac{1}{2}), 0\middle\vert R^{I-\frac{1}{2}}\middle\vert (I-\frac{1}{2}),0 \right\rangle \\
+& 2 (I+1) (2I+1) \left\langle (I+\frac{1}{2}), 0 \middle\vert R^{I+\frac{1}{2}}\middle\vert (I+\frac{1}{2}),0 \right\rangle +6 I \left\langle (I-\frac{1}{2}), 1 \middle\vert R^{I-\frac{1}{2}}\middle\vert (I-\frac{1}{2}),1 \right\rangle \\
 +& \frac{3 I \sqrt{2 I-1}}{\sqrt{I+1}} \biggl( \left\langle (I-\frac{1}{2}), 1\middle\vert R^{I-\frac{1}{2}}\middle\vert (I+\frac{1}{2}),1 \right\rangle + \left\langle (I+\frac{1}{2}), 1 \middle\vert R^{I-\frac{1}{2}}\middle\vert (I-\frac{1}{2}),1 \right\rangle \biggr) \\
  -& \frac{3 I (2 I-1)}{\sqrt{I (2 I+3)}}\biggl( \left\langle (I-\frac{1}{2}), 1 \middle\vert R^{I+\frac{1}{2}}\middle\vert (I+\frac{1}{2}),1 \right\rangle +  \left\langle (I+\frac{1}{2}), 1 \middle\vert R^{I+\frac{1}{2}}\middle\vert (I-\frac{1}{2}),1 \right\rangle \biggr) \\
  +& \frac{3 I^2 (3+2 I)}{I+1} \left\langle (I+\frac{1}{2}), 1 \middle\vert R^{I-\frac{1}{2}}\middle\vert (I+\frac{1}{2}),1 \right\rangle  +3 \bigl(2 I^2+I+1 \bigr) \left\langle (I-\frac{1}{2}), 1 \middle\vert R^{I+\frac{1}{2}}\middle\vert (I-\frac{1}{2}),1 \right\rangle   \\
  +& \frac{6 \bigl(2 I^2 +5 I+1\bigr)}{2 I+3} \left\langle (I+\frac{1}{2}), 1 \middle\vert R^{I+\frac{1}{2}}\middle\vert (I+\frac{1}{2}),1 \right\rangle  \Biggr)
\end{split}
\end{equation}
For the target spin $I=1/2$ and spin projection $M$:
\begin{equation}\label{bM12}
\begin{split}
B^{\prime}_{1/2M}=& - \frac{iM}{4 k  } \biggl( \left\langle 0, 0\middle\vert R^{0}\middle\vert 0,0 \right\rangle - \left\langle 1, 0\middle\vert R^{1}\middle\vert 1,0 \right\rangle
+3  \left\langle 0, 1\middle\vert R^{1}\middle\vert 0,1 \right\rangle- \left\langle 1, 1\middle\vert R^{0}\middle\vert 1,1 \right\rangle \biggr)
\end{split}
\end{equation}
For the target spin $I=1$ :
\begin{equation}\label{bM10}
\begin{split}
B^{\prime}_{1M}=& \frac{iM}{120 k  } \Biggl(-20\left\langle \frac{1}{2}, 0 \middle\vert R^{\frac{1}{2}}\middle\vert \frac{1}{2},0 \right\rangle +20\left\langle \frac{3}{2}, 0 \middle\vert R^{\frac{3}{2}}\middle\vert \frac{3}{2},0 \right\rangle -20\left\langle \frac{1}{2}, 1 \middle\vert R^{\frac{1}{2}}\middle\vert \frac{1}{2},1 \right\rangle \\
 -& 5 \sqrt{2}\left\langle \frac{1}{2}, 1 \middle\vert R^{\frac{1}{2}}\middle\vert \frac{3}{2},1 \right\rangle -40\left\langle \frac{1}{2}, 1 \middle\vert R^{\frac{3}{2}}\middle\vert \frac{1}{2},1 \right\rangle +2 \sqrt{5}\left\langle \frac{1}{2}, 1 \middle\vert R^{\frac{3}{2}}\middle\vert \frac{3}{2},1 \right\rangle \\
 -& 5 \sqrt{2}\left\langle \frac{3}{2}, 1 \middle\vert R^{\frac{1}{2}}\middle\vert \frac{1}{2},1 \right\rangle +20\left\langle \frac{3}{2}, 1 \middle\vert R^{\frac{1}{2}}\middle\vert \frac{3}{2},1 \right\rangle +2 \sqrt{5}\left\langle \frac{3}{2}, 1 \middle\vert R^{\frac{3}{2}}\middle\vert \frac{1}{2},1 \right\rangle \\
 +& 4\left\langle \frac{3}{2}, 1 \middle\vert R^{\frac{3}{2}}\middle\vert \frac{3}{2},1 \right\rangle \Biggr)
\end{split}
\end{equation}
For the target spin $I>1$:
\begin{equation}\label{bMany}
\begin{split}
B^{\prime}_{IM}=&\frac{iM}{4 (2 I+1)^2 k  } \\
&\Biggl(2 (2I+1) \biggl( \left\langle (I+\frac{1}{2}), 0 \middle\vert R^{I+\frac{1}{2}}\middle\vert (I+\frac{1}{2}),0 \right\rangle - \left\langle (I-\frac{1}{2}), 0\middle\vert R^{I-\frac{1}{2}}\middle\vert (I-\frac{1}{2}),0 \right\rangle \biggr) \\
-& \frac{6   \bigl(2 I^2+2 I-1-2 M^2\bigr)}{ (2 I-1)  } \left\langle (I-\frac{1}{2}), 1 \middle\vert R^{I-\frac{1}{2}}\middle\vert (I-\frac{1}{2}),1 \right\rangle  \\
-&\frac{3    \bigl(2 I^2+2 I-1-2 M^2\bigr)}{\sqrt{(I+1) (2 I-1)}} \biggl( \left\langle (I-\frac{1}{2}), 1\middle\vert R^{I-\frac{1}{2}}\middle\vert (I+\frac{1}{2}),1 \right\rangle + \left\langle (I+\frac{1}{2}), 1 \middle\vert R^{I-\frac{1}{2}}\middle\vert (I-\frac{1}{2}),1 \right\rangle \biggr) \\
-&\frac{3   \bigl(1+I+I^2+M^2\bigr)}{ I  } \left\langle (I-\frac{1}{2}), 1 \middle\vert R^{I+\frac{1}{2}}\middle\vert (I-\frac{1}{2}),1 \right\rangle  \\
+&\frac{3   \bigl(2 I^2+2 I-1-2 M^2\bigr)}{\sqrt{I (2 I+3)} } \biggl( \left\langle (I-\frac{1}{2}), 1 \middle\vert R^{I+\frac{1}{2}}\middle\vert (I+\frac{1}{2}),1 \right\rangle  + \left\langle (I+\frac{1}{2}), 1 \middle\vert R^{I+\frac{1}{2}}\middle\vert (I-\frac{1}{2}),1 \right\rangle \biggr) \\
+&\frac{3   \bigl(1+I+I^2+M^2\bigr) }{(I+1)  } \left\langle (I+\frac{1}{2}), 1 \middle\vert R^{I-\frac{1}{2}}\middle\vert (I+\frac{1}{2}),1 \right\rangle  \\
-&\frac{6   \bigl(2 I^2+2 I-1-2 M^2\bigr) }{ (2 I+3) } \left\langle (I+\frac{1}{2}), 1 \middle\vert R^{I+\frac{1}{2}}\middle\vert (I+\frac{1}{2}),1 \right\rangle  \Biggr) .
\end{split}
\end{equation}
For the target spin $I>1$  and $M=I$:
\begin{equation}\label{bMI}
\begin{split}
B^{\prime}_{II}=&\frac{i}{4 (2 I+1)^2 k  } \\
&\Biggl(  2 I (2 I+1)\biggl( \left\langle (I+\frac{1}{2}), 0 \middle\vert R^{I+\frac{1}{2}}\middle\vert (I+\frac{1}{2}),0 \right\rangle  - \left\langle (I-\frac{1}{2}), 0\middle\vert R^{I-\frac{1}{2}}\middle\vert (I-\frac{1}{2}),0 \right\rangle \biggr) \\
 +&3 \bigl(2 I^2+I+1 \bigr)\biggl( \frac{ I  }{I+1} \left\langle (I+\frac{1}{2}), 1 \middle\vert R^{I-\frac{1}{2}}\middle\vert (I+\frac{1}{2}),1 \right\rangle - \left\langle (I-\frac{1}{2}), 1 \middle\vert R^{I+\frac{1}{2}}\middle\vert (I-\frac{1}{2}),1 \right\rangle \biggr) \\
 -& \frac{3 I \sqrt{2 I-1} }{\sqrt{I+1}} \biggl( \left\langle (I-\frac{1}{2}), 1\middle\vert R^{I-\frac{1}{2}}\middle\vert (I+\frac{1}{2}),1 \right\rangle + \left\langle (I+\frac{1}{2}), 1 \middle\vert R^{I-\frac{1}{2}}\middle\vert (I-\frac{1}{2}),1 \right\rangle  \biggr) \\
  +& \frac{3 \sqrt{I} (2 I-1)}{\sqrt{2 I+3}} \biggl( \left\langle (I-\frac{1}{2}), 1 \middle\vert R^{I+\frac{1}{2}}\middle\vert (I+\frac{1}{2}),1 \right\rangle + \left\langle (I+\frac{1}{2}), 1 \middle\vert R^{I+\frac{1}{2}}\middle\vert (I-\frac{1}{2}),1 \right\rangle  \biggr) \\
     +& \frac{6 I (2 I-1)}{2 I+3} \left\langle (I+\frac{1}{2}), 1 \middle\vert R^{I+\frac{1}{2}}\middle\vert (I+\frac{1}{2}),1 \right\rangle  - 6 I \left\langle (I-\frac{1}{2}), 1 \middle\vert R^{I-\frac{1}{2}}\middle\vert (I-\frac{1}{2}),1 \right\rangle  \Biggr)
\end{split}
\end{equation}
For the target spin $I=1/2$ and its projections on axis $z$ $M=\pm 1/2$:
\begin{equation}\label{cM12}
\begin{split}
C^{\prime}_{1/2}= - \frac{i}{8 k  }& \biggl( \left\langle 0, 0\middle\vert R^{0}\middle\vert 1,1 \right\rangle
-\sqrt{3} \left\langle 1, 0\middle\vert R^{1}\middle\vert 0,1 \right\rangle
+\sqrt{6} \left\langle 1, 0\middle\vert R^{1}\middle\vert 1,1 \right\rangle \\
+& \left\langle 1, 1\middle\vert R^{0}\middle\vert 0,0 \right\rangle
- \sqrt{3} \left\langle 0, 1\middle\vert R^{1}\middle\vert 1,0 \right\rangle
+\sqrt{6} \left\langle 1, 1\middle\vert R^{1}\middle\vert 1,0 \right\rangle
\biggr)
\end{split}
\end{equation}
For the target spin $I>1/2$ :
\begin{equation}\label{cMany}
\begin{split}
C^{\prime}_{IM}=& -\frac{i \sqrt{3}}{8 (2 I+1)^{3/2} k  } \\
 &\Bigl(-\frac{4 \bigl(I^2-M^2\bigr)}{\sqrt{2 I-1}} \bigl( \left\langle (I-\frac{1}{2}), 0\middle\vert R^{I-\frac{1}{2}}\middle\vert (I-\frac{1}{2}),1 \right\rangle + \left\langle (I-\frac{1}{2}),1 \middle\vert R^{I-\frac{1}{2}}\middle\vert (I-\frac{1}{2}), 0 \right\rangle     \bigr) \\
+&\frac{2 (I^2+I+M^2)}{\sqrt{I+1}} \bigl( \left\langle (I-\frac{1}{2}), 0\middle\vert R^{I-\frac{1}{2}}\middle\vert (I+\frac{1}{2}),1 \right\rangle +\left\langle (I+\frac{1}{2}), 1\middle\vert R^{I-\frac{1}{2}}\middle\vert (I-\frac{1}{2}),0 \right\rangle   \bigr) \\
-&\frac{2 (I^2+I+M^2)}{\sqrt{I}} \bigl( \left\langle (I+\frac{1}{2}), 0\middle\vert R^{I+\frac{1}{2}}\middle\vert (I-\frac{1}{2}),1 \right\rangle   +\left\langle (I-\frac{1}{2}), 1\middle\vert R^{I+\frac{1}{2}}\middle\vert (I+\frac{1}{2}),0 \right\rangle \bigr) \\
+&\frac{4 ((I+1)^2-M^2) }{\sqrt{2 I+3}} \bigl( \left\langle (I+\frac{1}{2}), 0\middle\vert R^{I+\frac{1}{2}}\middle\vert (I+\frac{1}{2}),1 \right\rangle + \left\langle (I+\frac{1}{2}), 1\middle\vert R^{I+\frac{1}{2}}\middle\vert (I+\frac{1}{2}),0 \right\rangle      \bigr)    \Bigr) .
\end{split}
\end{equation}
Then, for the target spin $I>1/2$  and $M=I$:
\begin{equation}\label{cMI}
\begin{split}
C^{\prime}_{II}=&\frac{i \sqrt{3}}{4 \sqrt{2 I+1} k  } \Biggl( \frac{I}{\sqrt{I+1}} \bigl( \left\langle (I-\frac{1}{2}), 0\middle\vert R^{I-\frac{1}{2}}\middle\vert (I+\frac{1}{2}),1 \right\rangle
+ \left\langle (I+\frac{1}{2}), 1\middle\vert R^{I-\frac{1}{2}}\middle\vert (I-\frac{1}{2}),0 \right\rangle \bigr)  \\
-& \sqrt{I} \bigl( \left\langle (I+\frac{1}{2}), 0\middle\vert R^{I+\frac{1}{2}}\middle\vert (I-\frac{1}{2}),1 \right\rangle
+ \left\langle (I-\frac{1}{2}), 1\middle\vert R^{I+\frac{1}{2}}\middle\vert (I+\frac{1}{2}),0 \right\rangle \bigr) \\
+& \frac{2}{\sqrt{2 I+3}}\bigl(  \left\langle (I+\frac{1}{2}), 0\middle\vert R^{I+\frac{1}{2}}\middle\vert (I+\frac{1}{2}),1 \right\rangle
+  \left\langle (I+\frac{1}{2}), 1\middle\vert R^{I+\frac{1}{2}}\middle\vert (I+\frac{1}{2}),0 \right\rangle \bigr) \Biggr)
\end{split}
\end{equation}
The general expression for $D^{\prime}$ is:
\begin{equation}\label{dM}
\begin{split}
D^{\prime}_M=&\frac{\sqrt{3} M}{4 \sqrt{I (I+1) (2 I+1)} k  }\\
 & \Biggl(\sqrt{I+1} \Bigl(\left\langle (I+\frac{1}{2}), 0\middle\vert R^{I+\frac{1}{2}}\middle\vert (I-\frac{1}{2}),1 \right\rangle-\left\langle (I-\frac{1}{2}), 1\middle\vert R^{I+\frac{1}{2}}\middle\vert (I+\frac{1}{2}),0 \right\rangle\Bigr) \\
 + &\sqrt{I} \Bigl(\left\langle (I-\frac{1}{2}), 0\middle\vert R^{I-\frac{1}{2}}\middle\vert (I+\frac{1}{2}),1 \right\rangle-\left\langle (I+\frac{1}{2}), 1\middle\vert R^{I-\frac{1}{2}}\middle\vert (I-\frac{1}{2}),0 \right\rangle\Bigr)\Biggr) ,
\end{split}
\end{equation}
where $M$ is a projection of spin $I$ on axis $z$.

It should be noted that for the fixed value $M$ some of the above coefficients contain a mixture from different types of the target polarizations. It can be clear seen from the case of $M=I$: the 100 \%  population of this level contributes to all possible tensor polarizations $P_q$ up to the rank $q=2I$ (see appendix \ref{tensor}).

\section{$d$-wave contributions}
\label{dwave}

To estimate  contributions from $d$-waves let us consider  $E^{\prime}$ and $F^{\prime}$ coefficient
obtained  the scattering amplitude related to tensor polarization $t_{20}$ with $d$-waves
\begin{eqnarray}\label{20Amp}
 \nonumber
 f_{20} (\theta ,\phi ,x,y) & =&\frac{i\pi }{2k}\sqrt{\frac{5}{2I+1}}t_{20} \sum_{JMll^{\prime}  S  S^{\prime} m}Y_{L m}(\theta ,\phi)N(x,y)
 \left\langle  I M 20\middle\vert IM \right\rangle \\
     &\times&
   \nonumber
  	\left\langle S^{\prime} l^{\prime} \alpha^{\prime} \middle\vert R^J \middle\vert S l \alpha\right\rangle
	(-1)^{J+S^{\prime}+l^{\prime}+l}(2J+1)\sqrt{\frac{(2l+1)(2l^{\prime}+1)}{4\pi (2S+1)}} \\
	 &\times&
\left\langle l 0 l^{\prime} 0 \middle\vert L 0 \right\rangle
 \begin{Bmatrix}
   l^{\prime} & l & L \\
   S & S^{\prime} & J
  \end{Bmatrix}.
 \end{eqnarray}

The  expression for $E^{\prime}$ for the tensor polarized $\widetilde{\tau}_{20}$ target with $I=7/2$ is:
\begin{equation}\label{Ep72}
\begin{split}
E^{\prime}_{7/2}=&-\frac{i}{512 \sqrt{2} k  } \Bigl(5 \sqrt{\frac{21}{2}} \left\langle 3, 1 \middle\vert
R^{3}\middle\vert 3,1 \right\rangle -15 \sqrt{\frac{3}{14}}
\left\langle 3, 1 \middle\vert
R^{4}\middle\vert 3,1 \right\rangle-11
\sqrt{\frac{7}{6}} \left\langle 4, 1 \middle\vert
R^{3}\middle\vert 4,1 \right\rangle \\
- &3 \sqrt{\frac{7}{2}} (\left\langle 3, 1 \middle\vert
R^{3}\middle\vert 4,1 \right\rangle+\left\langle 4, 1 \middle\vert
R^{3}\middle\vert 3,1 \right\rangle)+9 \sqrt{\frac{3}{10}}
(\left\langle 3, 1 \middle\vert
R^{4}\middle\vert 4,1 \right\rangle+\left\langle 4, 1 \middle\vert
R^{4}\middle\vert 3,1 \right\rangle) \\
+& \frac{33}{5} \sqrt{\frac{21}{2}}
\left\langle 4, 1 \middle\vert R^{4}\middle\vert 4,1 \right\rangle+\frac{55}{\sqrt{42}}
\left\langle 3, 2 \middle\vert R^{3}\middle\vert 3,2 \right\rangle+\frac{225}{7} \sqrt{\frac{3}{14}}\left\langle 3, 2 \middle\vert R^{4}\middle\vert 3,2 \right\rangle \\
+& 5 \sqrt{14} (\left\langle 3, 0 \middle\vert R^{3}\middle\vert 3,2 \right\rangle+\left\langle 3, 2 \middle\vert R^{3}\middle\vert 3,0 \right\rangle)
+3 \sqrt{6} (\left\langle 4, 0 \middle\vert R^{4}\middle\vert 3,2 \right\rangle+\left\langle 3, 2 \middle\vert R^{4}\middle\vert 4,0 \right\rangle) \\
-& \sqrt{42}
(\left\langle 3, 0 \middle\vert R^{3}\middle\vert 4,2 \right\rangle+\left\langle 4, 2 \middle\vert R^{3}\middle\vert 3,0 \right\rangle)-\frac{15 }{\sqrt{14}}
(\left\langle 3, 2 \middle\vert R^{3}\middle\vert 4,2 \right\rangle+\left\langle 4, 2 \middle\vert R^{3}\middle\vert 3,2 \right\rangle) \\
-&\frac{3}{7} \sqrt{\frac{33}{14}}
(\left\langle 3, 2 \middle\vert R^{4}\middle\vert 4,2 \right\rangle+\left\langle 4, 2 \middle\vert R^{4}\middle\vert 3,2 \right\rangle)+3 \sqrt{66}
(\left\langle 4, 0 \middle\vert R^{4}\middle\vert 4,2 \right\rangle+\left\langle 4, 2 \middle\vert R^{4}\middle\vert 4,0 \right\rangle) \\
-&\frac{11}{\sqrt{42}}
\left\langle 4, 2 \middle\vert R^{3}\middle\vert 4,2 \right\rangle+\frac{195}{7} \sqrt{\frac{3}{14}} \left\langle 4, 2 \middle\vert R^{4}\middle\vert 4,2 \right\rangle\Bigr) .
\end{split}
\end{equation}
One can see that   $E^{\prime}$ coefficient is equal to zero at $s$-resonances and depends on $p$-wave and $d$-wave resonances. However, contributions from $d$-wave resonances  are suppressed in low energy  region by a factor $(kR)$ in   compare to $p$-wave ones (where $R$ is a nuclear radius). Moreover, they are behave in the vicinity of $p$-wave resonances as a flat (energy independent) background. Therefore $d$-wave contributions are negligible in the vicinity of low energy $p$-wave resonances.

The  expression for $F^{\prime}$ for the tensor polarized  target with $I=7/2$ is:
\begin{equation}\label{Fp72}
\begin{split}
F^{\prime}_{7/2}=&-\frac{i}{128 k  } \Bigl(\sqrt{\frac{21}{2}} (\left\langle 3, 0 \middle\vert
R^{3}\middle\vert 3,1 \right\rangle+\left\langle 3, 1 \middle\vert
R^{3}\middle\vert 3,0 \right\rangle)-\frac{3 }{\sqrt{2}} (\left\langle 4, 0 \middle\vert
R^{4}\middle\vert 3,1 \right\rangle)+\left\langle 3, 1 \middle\vert
R^{4}\middle\vert 4,0 \right\rangle) \\
+&\sqrt{\frac{7}{2}} (\left\langle 3, 0 \middle\vert
R^{3}\middle\vert 4,1 \right\rangle+\left\langle 4, 1 \middle\vert
R^{3}\middle\vert 3,0 \right\rangle)-3 \sqrt{\frac{7}{10}} (\left\langle 4, 0 \middle\vert
R^{4}\middle\vert 4,1 \right\rangle+\left\langle 4, 1 \middle\vert
R^{4}\middle\vert 4,0 \right\rangle) \\
+&\frac{15}{\sqrt{14}} (\left\langle 3, 1 \middle\vert
R^{3}\middle\vert 3,2 \right\rangle+\left\langle 3, 2 \middle\vert
R^{3}\middle\vert 3,1 \right\rangle)-10 \sqrt{\frac{2}{21}} (\left\langle 4, 1 \middle\vert
R^{3}\middle\vert 3,2 \right\rangle+\left\langle 3, 2 \middle\vert
R^{3}\middle\vert 4,1 \right\rangle) \\
+&\frac{81}{7 \sqrt{14}} (\left\langle 3, 1 \middle\vert
R^{4}\middle\vert 3,2 \right\rangle+\left\langle 3, 2 \middle\vert
R^{4}\middle\vert 3,1 \right\rangle)+\frac{9}{7} \sqrt{\frac{2}{5}} (\left\langle 4, 1 \middle\vert
R^{4}\middle\vert 3,2 \right\rangle+\left\langle 3, 2 \middle\vert
R^{4}\middle\vert 4,1 \right\rangle) \\
-&\sqrt{\frac{6}{7}} (\left\langle 3, 1 \middle\vert
R^{3}\middle\vert 4,2 \right\rangle+\left\langle 4, 2 \middle\vert
R^{3}\middle\vert 3,1 \right\rangle)+\frac{5}{\sqrt{14}} (\left\langle 4, 1 \middle\vert
R^{3}\middle\vert 4,2 \right\rangle+\left\langle 4, 2 \middle\vert
R^{3}\middle\vert 4,1 \right\rangle) \\
+&\frac{12}{7} \sqrt{\frac{22}{7}} (\left\langle 3, 1 \middle\vert
R^{4}\middle\vert 4,2 \right\rangle+\left\langle 4, 2 \middle\vert
R^{4}\middle\vert 3,1 \right\rangle)-\frac{39}{7} \sqrt{\frac{11}{10}} (\left\langle 4, 1 \middle\vert
R^{4}\middle\vert 4,2 \right\rangle+\left\langle 4, 2 \middle\vert
R^{4}\middle\vert 4,1 \right\rangle)\Bigr) .
\end{split}
\end{equation}
we can see that   $F^{\prime}$ coefficient defined by P-odd mixtures of $s$-wave and $p$-wave resonances, and $p$-wave and $d$-wave resonances. Therefore, contributions from  $d$-wave resonances are suppressed in  low energy region by a factor $(kR)^2$, and can be  neglected.

\section{None-zero
contributions in a spherical tensor expansion of the amplitude in eq.(\ref{tensAmpNeutr}).}
\label{Fappendix}

The $f_0$ for $q=0$:
\begin{equation}\label{F0}
\begin{split}
f_{0}=& \frac{i}{64  k  } \Biggl( 2 \Bigl(  7\left\langle 3, 0 \middle\vert R^{3}\middle\vert 3,0 \right\rangle +9\left\langle 4, 0 \middle\vert R^{4}\middle\vert 4,0 \right\rangle +7\left\langle 3, 1 \middle\vert R^{3}\middle\vert 3,1 \right\rangle \\
&+9\left\langle 3, 1 \middle\vert R^{4}\middle\vert 3,1 \right\rangle
+7\left\langle 4, 1 \middle\vert R^{3}\middle\vert 4,1 \right\rangle +9\left\langle 4, 1 \middle\vert R^{4}\middle\vert 4,1 \right\rangle \Bigr) \\
&+\Bigl( 7(\left\langle 3, 0 \middle\vert R^{3}\middle\vert 3,1 \right\rangle +\left\langle 3, 1 \middle\vert R^{3}\middle\vert 3,0 \right\rangle  )
 -7 \sqrt{3}(\left\langle 3, 0 \middle\vert R^{3}\middle\vert 4,1 \right\rangle +\left\langle 4, 1 \middle\vert R^{3}\middle\vert 3,0 \right\rangle ) \\
 &+3 \sqrt{21}(\left\langle 4, 0 \middle\vert R^{4}\middle\vert 3,1 \right\rangle +\left\langle 3, 1 \middle\vert R^{4}\middle\vert 4,0 \right\rangle   ) -3 \sqrt{15}(\left\langle 4, 0 \middle\vert R^{4}\middle\vert 4,1 \right\rangle
  +\left\langle 4, 1 \middle\vert R^{4}\middle\vert 4,0 \right\rangle ) \Bigr)\\
 & \times (\cos (\beta ) \cos (\theta )+\sin (\alpha ) \sin (\beta ) \sin (\theta )) \Biggr) ,
\end{split}
\end{equation}
or
\begin{equation}\label{F0vec}
\begin{split}
f_{0}=& \frac{i}{64  k  } \Biggl( 2 \Bigl( 7\left\langle 3, 0 \middle\vert R^{3}\middle\vert 3,0 \right\rangle +9\left\langle 4, 0 \middle\vert R^{4}\middle\vert 4,0 \right\rangle +7\left\langle 3, 1 \middle\vert R^{3}\middle\vert 3,1 \right\rangle \\
&+9\left\langle 3, 1 \middle\vert R^{4}\middle\vert 3,1 \right\rangle
+7\left\langle 4, 1 \middle\vert R^{3}\middle\vert 4,1 \right\rangle +9\left\langle 4, 1 \middle\vert R^{4}\middle\vert 4,1 \right\rangle \Bigr) \\
&+\Bigl( 7(\left\langle 3, 0 \middle\vert R^{3}\middle\vert 3,1 \right\rangle +\left\langle 3, 1 \middle\vert R^{3}\middle\vert 3,0 \right\rangle  )
 -7 \sqrt{3}(\left\langle 3, 0 \middle\vert R^{3}\middle\vert 4,1 \right\rangle +\left\langle 4, 1 \middle\vert R^{3}\middle\vert 3,0 \right\rangle ) \\
 &+3 \sqrt{21}(\left\langle 4, 0 \middle\vert R^{4}\middle\vert 3,1 \right\rangle +\left\langle 3, 1 \middle\vert R^{4}\middle\vert 4,0 \right\rangle   ) -3 \sqrt{15}(\left\langle 4, 0 \middle\vert R^{4}\middle\vert 4,1 \right\rangle
  +\left\langle 4, 1 \middle\vert R^{4}\middle\vert 4,0 \right\rangle ) \Bigr)\\
 & \times (\vec{\sigma} \cdot \vec{k}) \Biggr) .
\end{split}
\end{equation}
The $f_1$ for $q=1$:
\begin{equation}\label{F1}
\begin{split}
f_{1}=& -\frac{i}{32 k  }  \Biggl( 7\left\langle 3, 0 \middle\vert R^{3}\middle\vert 3,0 \right\rangle - 7\left\langle 4, 0 \middle\vert R^{4}\middle\vert 4,0 \right\rangle \\
 &+\frac{21}{4} \left\langle 3, 1 \middle\vert R^{3}\middle\vert 3,1 \right\rangle +\frac{7}{4} \sqrt{3}(\left\langle 3, 1 \middle\vert R^{3}\middle\vert 4,1 \right\rangle +\left\langle 4, 1 \middle\vert R^{3}\middle\vert 3,1 \right\rangle )  \\
 &-\frac{9}{20} \sqrt{35}(\left\langle 3, 1 \middle\vert R^{4}\middle\vert 4,1 \right\rangle + \left\langle 4, 1 \middle\vert R^{4}\middle\vert 3,1 \right\rangle  )  -\frac{91}{12} \left\langle 4, 1 \middle\vert R^{3}\middle\vert 4,1 \right\rangle \\
 & +\frac{39}{4} \left\langle 3, 1 \middle\vert R^{4}\middle\vert 3,1 \right\rangle-\frac{63}{20} \left\langle 4, 1 \middle\vert R^{4}\middle\vert 4,1 \right\rangle  \Biggr)\cos (\beta ) \\
  &+\frac{1}{16 k  }\Biggl( \frac{7}{\sqrt{3}}(\left\langle 3, 0 \middle\vert R^{3}\middle\vert 4,1 \right\rangle -\left\langle 4, 1 \middle\vert R^{3}\middle\vert 3,0 \right\rangle  ) + \sqrt{21}(\left\langle 4, 0 \middle\vert R^{4}\middle\vert 3,1 \right\rangle -\left\langle 3, 1 \middle\vert R^{4}\middle\vert 4,0 \right\rangle ) \Biggr) \\
 &\times \cos (\alpha ) \sin (\beta ) \sin (\theta ) \\
   &-\frac{i }{64 k  } \Biggl( 21(\left\langle 3, 0 \middle\vert R^{3}\middle\vert 3,1 \right\rangle + \left\langle 3, 1 \middle\vert R^{3}\middle\vert 3,0 \right\rangle ) -\frac{7}{\sqrt{3}}(\left\langle 3, 0 \middle\vert R^{3}\middle\vert 4,1 \right\rangle  +\left\langle 4, 1 \middle\vert R^{3}\middle\vert 3,0 \right\rangle  ) \\
  &+\sqrt{21}(\left\langle 4, 0 \middle\vert R^{4}\middle\vert 3,1 \right\rangle +\left\langle 3, 1 \middle\vert R^{4}\middle\vert 4,0 \right\rangle ) +7 \sqrt{15}(\left\langle 4, 0 \middle\vert R^{4}\middle\vert 4,1 \right\rangle
    +\left\langle 4, 1 \middle\vert R^{4}\middle\vert 4,0 \right\rangle ) \Biggr)\cos (\theta ) \\
      &-\frac{3i }{128 k  } \Biggl( 7\left\langle 3, 1 \middle\vert R^{3}\middle\vert 3,1 \right\rangle -7 \sqrt{3}(\left\langle 3, 1 \middle\vert R^{3}\middle\vert 4,1 \right\rangle +\left\langle 4, 1 \middle\vert R^{3}\middle\vert 3,1 \right\rangle  )  \\
   & +\frac{77}{9\sqrt{3}} \left\langle 4, 1 \middle\vert R^{3}\middle\vert 4,1 \right\rangle-3\left\langle 3, 1 \middle\vert R^{4}\middle\vert 3,1 \right\rangle +\frac{9}{5} \sqrt{35}(\left\langle 3, 1 \middle\vert R^{4}\middle\vert 4,1 \right\rangle + \left\langle 4, 1 \middle\vert R^{4}\middle\vert 3,1 \right\rangle )
    \\
    &  -\frac{77}{5} \left\langle 4, 1 \middle\vert R^{4}\middle\vert 4,1 \right\rangle \Biggr) (\cos (\beta ) \cos (\theta )+\sin (\alpha ) \sin (\beta ) \sin (\theta ))\cos (\theta ) ,
\end{split}
\end{equation}
or
\begin{equation}\label{F1vec}
\begin{split}
f_{1}=& -\frac{i}{32 k  }  \Biggl( 7\left\langle 3, 0 \middle\vert R^{3}\middle\vert 3,0 \right\rangle - 7\left\langle 4, 0 \middle\vert R^{4}\middle\vert 4,0 \right\rangle \\
 &+\frac{21}{4} \left\langle 3, 1 \middle\vert R^{3}\middle\vert 3,1 \right\rangle +\frac{7}{4} \sqrt{3}(\left\langle 3, 1 \middle\vert R^{3}\middle\vert 4,1 \right\rangle +\left\langle 4, 1 \middle\vert R^{3}\middle\vert 3,1 \right\rangle )  \\
 &-\frac{9}{20} \sqrt{35}(\left\langle 3, 1 \middle\vert R^{4}\middle\vert 4,1 \right\rangle + \left\langle 4, 1 \middle\vert R^{4}\middle\vert 3,1 \right\rangle  )  -\frac{91}{12} \left\langle 4, 1 \middle\vert R^{3}\middle\vert 4,1 \right\rangle \\
 & +\frac{39}{4} \left\langle 3, 1 \middle\vert R^{4}\middle\vert 3,1 \right\rangle-\frac{63}{20} \left\langle 4, 1 \middle\vert R^{4}\middle\vert 4,1 \right\rangle  \Biggr) (\vec{\sigma}\cdot \vec{I}) \\
 &+\frac{1}{16 k  }\Biggl( \frac{7}{\sqrt{3}}(\left\langle 3, 0 \middle\vert R^{3}\middle\vert 4,1 \right\rangle -\left\langle 4, 1 \middle\vert R^{3}\middle\vert 3,0 \right\rangle  ) +\sqrt{21}(\left\langle 4, 0 \middle\vert R^{4}\middle\vert 3,1 \right\rangle -\left\langle 3, 1 \middle\vert R^{4}\middle\vert 4,0 \right\rangle ) \Biggr) \\
 &\times (\vec{\sigma}\cdot [\vec{k}\times \vec{I}]) \\
   &-\frac{i }{64 k  } \Biggl( 21(\left\langle 3, 0 \middle\vert R^{3}\middle\vert 3,1 \right\rangle + \left\langle 3, 1 \middle\vert R^{3}\middle\vert 3,0 \right\rangle ) -\frac{7}{\sqrt{3}}(\left\langle 3, 0 \middle\vert R^{3}\middle\vert 4,1 \right\rangle  +\left\langle 4, 1 \middle\vert R^{3}\middle\vert 3,0 \right\rangle  ) \\
  &+\sqrt{21}(\left\langle 4, 0 \middle\vert R^{4}\middle\vert 3,1 \right\rangle +\left\langle 3, 1 \middle\vert R^{4}\middle\vert 4,0 \right\rangle ) +7 \sqrt{15}(\left\langle 4, 0 \middle\vert R^{4}\middle\vert 4,1 \right\rangle
    +\left\langle 4, 1 \middle\vert R^{4}\middle\vert 4,0 \right\rangle ) \Biggr)
    (\vec{k} \cdot \vec{I}) \\
  &-\frac{3i }{128 k  } \Biggl( 7\left\langle 3, 1 \middle\vert R^{3}\middle\vert 3,1 \right\rangle -7 \sqrt{3}(\left\langle 3, 1 \middle\vert R^{3}\middle\vert 4,1 \right\rangle +\left\langle 4, 1 \middle\vert R^{3}\middle\vert 3,1 \right\rangle  )  \\
   & +\frac{77}{9\sqrt{3}} \left\langle 4, 1 \middle\vert R^{3}\middle\vert 4,1 \right\rangle-3\left\langle 3, 1 \middle\vert R^{4}\middle\vert 3,1 \right\rangle +\frac{9}{5} \sqrt{35}(\left\langle 3, 1 \middle\vert R^{4}\middle\vert 4,1 \right\rangle + \left\langle 4, 1 \middle\vert R^{4}\middle\vert 3,1 \right\rangle )
    \\
    &  -\frac{77}{5} \left\langle 4, 1 \middle\vert R^{4}\middle\vert 4,1 \right\rangle \Biggr) (\vec{\sigma} \cdot \vec{k})(\vec{k} \cdot \vec{I}) .
\end{split}
\end{equation}
The $f_2$ for $q=2$:
\begin{equation}\label{F2}
\begin{split}
f_{2}=& \frac{3i}{320 k  }\Biggl( 35(\left\langle 3, 0 \middle\vert R^{3}\middle\vert 3,1 \right\rangle +\left\langle 3, 1 \middle\vert R^{3}\middle\vert 3,0 \right\rangle  ) +\frac{35}{ \sqrt{3}}(\left\langle 3, 0 \middle\vert R^{3}\middle\vert 4,1 \right\rangle +\left\langle 4, 1 \middle\vert R^{3}\middle\vert 3,0 \right\rangle ) \\
 &-5 \sqrt{21}(\left\langle 4, 0 \middle\vert R^{4}\middle\vert 3,1 \right\rangle +\left\langle 3, 1 \middle\vert R^{4}\middle\vert 4,0 \right\rangle ) -7 \sqrt{15}(\left\langle 4, 0 \middle\vert R^{4}\middle\vert 4,1 \right\rangle
    +\left\langle 4, 1 \middle\vert R^{4}\middle\vert 4,0 \right\rangle ) \Biggr) \\
  &\times \Bigl(\cos (\beta ) \cos (\theta )-\frac{1}{3}( \cos (\beta ) \cos (\theta )+\sin (\alpha ) \sin (\beta ) \sin (\theta )) \Bigr) \\
    &+\frac{i 9 }{256 k  } \Biggl( \frac{35}{3}  \left\langle 3, 1 \middle\vert R^{3}\middle\vert 3,1 \right\rangle    - \frac{7}{\sqrt{3}} ( \left\langle 3, 1 \middle\vert R^{3}\middle\vert 4,1 \right\rangle +\left\langle 4, 1 \middle\vert R^{3}\middle\vert 3,1 \right\rangle  ) \\
   & -\frac{77}{9} \left\langle 4, 1 \middle\vert R^{3}\middle\vert 4,1 \right\rangle +\frac{3}{5} \sqrt{35}( \left\langle 3, 1 \middle\vert R^{4}\middle\vert 4,1 \right\rangle  + \left\langle 4, 1 \middle\vert R^{4}\middle\vert 3,1 \right\rangle )\\
   & -5\left\langle 3, 1 \middle\vert R^{4}\middle\vert 3,1 \right\rangle
   +\frac{77}{5} \left\langle 4, 1 \middle\vert R^{4}\middle\vert 4,1 \right\rangle \Bigr) ( \cos^2 ( \theta )-\frac{1}{3})\\
   &-\frac{3 }{32 \sqrt{5} k  } \Biggl( 7\sqrt{\frac{5}{3}} (\left\langle 3, 1 \middle\vert R^{3}\middle\vert 4,1 \right\rangle -\left\langle 4, 1 \middle\vert R^{3}\middle\vert 3,1 \right\rangle ) \\
    &-3 \sqrt{7}(\left\langle 3, 1 \middle\vert R^{4}\middle\vert 4,1 \right\rangle -\left\langle 4, 1 \middle\vert R^{4}\middle\vert 3,1 \right\rangle )
    \Biggr)  \sin (\beta ) \sin (\theta )\cos (\theta ) \cos (\alpha ) ,
\end{split}
\end{equation}
or
\begin{equation}\label{F2vec}
\begin{split}
f_{2}=& \frac{3i}{320 k  }\Biggl( 35(\left\langle 3, 0 \middle\vert R^{3}\middle\vert 3,1 \right\rangle +\left\langle 3, 1 \middle\vert R^{3}\middle\vert 3,0 \right\rangle  ) +\frac{35 }{\sqrt{3}}(\left\langle 3, 0 \middle\vert R^{3}\middle\vert 4,1 \right\rangle +\left\langle 4, 1 \middle\vert R^{3}\middle\vert 3,0 \right\rangle ) \\
 &-5\sqrt{21}(\left\langle 4, 0 \middle\vert R^{4}\middle\vert 3,1 \right\rangle +\left\langle 3, 1 \middle\vert R^{4}\middle\vert 4,0 \right\rangle ) -7 \sqrt{15}(\left\langle 4, 0 \middle\vert R^{4}\middle\vert 4,1 \right\rangle
    +\left\langle 4, 1 \middle\vert R^{4}\middle\vert 4,0 \right\rangle ) \Biggr) \\
  &\times \Bigl( (\vec{\sigma}\cdot \vec{I})(\vec{k}\cdot \vec{I})-\frac{1}{3}(\vec{\sigma}\cdot \vec{k})(\vec{I}\cdot \vec{I}) \Bigr) \\
  &+\frac{i 9 }{256 k  } \Biggl( \frac{35}{3}  \left\langle 3, 1 \middle\vert R^{3}\middle\vert 3,1 \right\rangle    - \frac{7}{\sqrt{3}} ( \left\langle 3, 1 \middle\vert R^{3}\middle\vert 4,1 \right\rangle +\left\langle 4, 1 \middle\vert R^{3}\middle\vert 3,1 \right\rangle  ) \\
   & -\frac{77}{9} \left\langle 4, 1 \middle\vert R^{3}\middle\vert 4,1 \right\rangle +\frac{3}{5} \sqrt{35}( \left\langle 3, 1 \middle\vert R^{4}\middle\vert 4,1 \right\rangle  + \left\langle 4, 1 \middle\vert R^{4}\middle\vert 3,1 \right\rangle )\\
   & -5\left\langle 3, 1 \middle\vert R^{4}\middle\vert 3,1 \right\rangle
   +\frac{77}{5} \left\langle 4, 1 \middle\vert R^{4}\middle\vert 4,1 \right\rangle \Bigr) \left( (\vec{k}\cdot \vec{I})(\vec{k}\cdot \vec{I})-\frac{1}{3}(\vec{k}\cdot \vec{k})(\vec{I}\cdot \vec{I})\right)\\
   &-\frac{3 }{32 \sqrt{5} k  } \Biggl( 7\sqrt{\frac{5}{3}} (\left\langle 3, 1 \middle\vert R^{3}\middle\vert 4,1 \right\rangle -\left\langle 4, 1 \middle\vert R^{3}\middle\vert 3,1 \right\rangle ) \\
    &-3 \sqrt{7}(\left\langle 3, 1 \middle\vert R^{4}\middle\vert 4,1 \right\rangle -\left\langle 4, 1 \middle\vert R^{4}\middle\vert 3,1 \right\rangle )
    \Biggr)  (\vec{\sigma}\cdot [\vec{k}\times \vec{I}])(\vec{k}\cdot \vec{I}) .
\end{split}
\end{equation}
The $f_3$ for $q=3$:
\begin{equation}\label{F3}
\begin{split}
f_{3}=& -\frac{3i}{512 k}   \Biggl(  7\left\langle 3, 1 \middle\vert R^{3}\middle\vert 3,1 \right\rangle +\frac{7}{\sqrt{3}}(\left\langle 3, 1 \middle\vert R^{3}\middle\vert 4,1 \right\rangle +\left\langle 4, 1 \middle\vert R^{3}\middle\vert 3,1 \right\rangle  ) \\
& +     \frac{7}{3}\left\langle 4, 1 \middle\vert R^{3}\middle\vert 4,1 \right\rangle
 -3\sqrt{\frac{7}{5}}(\left\langle 3, 1 \middle\vert R^{4}\middle\vert 4,1 \right\rangle +\left\langle 4, 1 \middle\vert R^{4}\middle\vert 3,1 \right\rangle  )  \\
     &-3 \left\langle 3, 1 \middle\vert R^{4}\middle\vert 3,1 \right\rangle  -\frac{21 }{5} \left\langle 4, 1 \middle\vert R^{4}\middle\vert 4,1 \right\rangle
      \Biggr) \\
       &\times \Bigl(\cos (\beta )+3 \cos (\beta ) \cos (2 \theta )-4 \cos (\theta ) \sin (\alpha ) \sin (\beta ) \sin (\theta )\Bigr)  ,
\end{split}
\end{equation}
or
\begin{equation}\label{F3vec}
\begin{split}
f_{3}=&\frac{9i}{256 k  }   \Biggl( 7\left\langle 3, 1 \middle\vert R^{3}\middle\vert 3,1 \right\rangle +\frac{7}{\sqrt{3}}(\left\langle 3, 1 \middle\vert R^{3}\middle\vert 4,1 \right\rangle +\left\langle 4, 1 \middle\vert R^{3}\middle\vert 3,1 \right\rangle  ) \\
& +     \frac{7}{3} \left\langle 4, 1 \middle\vert R^{3}\middle\vert 4,1 \right\rangle
 -3\sqrt{\frac{7}{5}}(\left\langle 3, 1 \middle\vert R^{4}\middle\vert 4,1 \right\rangle +\left\langle 4, 1 \middle\vert R^{4}\middle\vert 3,1 \right\rangle  )  \\
     &-3\left\langle 3, 1 \middle\vert R^{4}\middle\vert 3,1 \right\rangle  -\frac{21 }{5} \left\langle 4, 1 \middle\vert R^{4}\middle\vert 4,1 \right\rangle
      \Biggr) \\
       &\times \Bigl(  (\vec{\sigma} \cdot \vec{I})[(\vec{k} \cdot \vec{I})(\vec{k} \cdot \vec{I})-\frac{1}{3}(\vec{k} \cdot \vec{k})(\vec{I} \cdot \vec{I})]
        +\frac{2}{5}(\vec{k} \cdot \vec{I})[(\vec{\sigma} \cdot \vec{I})(\vec{k} \cdot \vec{I})\frac{1}{3}+(\vec{\sigma} \cdot \vec{k})(\vec{I} \cdot \vec{I})] \\
        &-\frac{4}{5}(\vec{k} \cdot \vec{I})[(\vec{\sigma} \cdot \vec{I})(\vec{k} \cdot \vec{I})-\frac{1}{3}(\vec{\sigma} \cdot \vec{k})(\vec{I} \cdot \vec{I})]\Bigr)  .
\end{split}
\end{equation}

\bibliography{TViolation,ParityViolation}
\end{document}